\RequirePackage[T1]{fontenc}
\documentclass[12pt]{article}
\pdfoutput=1 
\usepackage[height=8.85in,width=6.45in]{geometry}

\usepackage[utf8]{inputenc}
\usepackage{amsmath}
\usepackage{amssymb}
\usepackage{mathtools}
\numberwithin{equation}{section}
\usepackage{slashed}
\usepackage{braket}
\usepackage[svgnames]{xcolor}
\usepackage[colorlinks,linktocpage=true,citecolor=DarkGreen,linkcolor=FireBrick]{hyperref}
\usepackage{cite}
\usepackage{graphicx}
\usepackage{tikz}
\usepackage{tikz-cd}
\usepackage{times}
\usepackage[scaled]{couriers}
\usepackage{bm}
\usepackage{subfig}
\usepackage{mathrsfs}
\usepackage{amsthm}

\usepackage{xcolor}
\usepackage{mdframed}

\renewenvironment{figure}[1][]{
  \begin{originalfigure}[#1]
    \begin{mdframed}[linecolor=black!0,backgroundcolor=black!1]
}{
    \end{mdframed}
  \end{originalfigure}
}

\theoremstyle{plain}

\theoremstyle{definition}

\numberwithin{thm}{section}

\def\d{{\rm d}}
\def\i{{\mathsf i}}

\DeclareMathOperator{\tr}{tr}

\def\Hom{\mathop{\mathrm{Hom}}}
\usepackage{slashed}

\def\cE{{\cal E}}

\def\cL{{\cal L}}

\def\cN{{\cal N}}

\def\cS{{\cal S}}

\def\cV{{\cal V}}

\def\cZ{{\cal Z}}

\def\bA{{\mathbb A}}

\def\bC{{\mathbb C}}

\def\bH{{\mathbb H}}

\def\bM{{\mathbb M}}

\def\bR{{\mathbb R}}

\def\bZ{{\mathbb Z}}

\def\sA{{\mathsf A}}
\def\sB{{\mathsf B}}
\def\sC{{\mathsf C}}

\def\sF{{\mathsf F}}
\def\sG{{\mathsf G}}
\def\sH{{\mathsf H}}

\def\sa{{\mathsf a}}
\def\sb{{\mathsf b}}
\def\sc{{\mathsf c}}

\def\so{{\mathsf o}}

\def\su{{\mathsf u}}

\def\sw{{\mathsf w}}

\def\U{\mathrm{U}}
\def\SU{\mathrm{SU}}

\def\Sp{\mathrm{Sp}}

\def\Spin{\mathrm{Spin}}

\def\SL{\mathrm{SL}}
\def\Mp{\mathrm{Mp}}
\def\GL{\mathrm{GL}}

\def\su{\mathfrak{su}}

\def\so{\mathfrak{so}}

\def\spin{\mathrm{spin}}
\def\pin{\mathrm{pin}}

\def\beq#1\eeq{\begin{align}#1\end{align}}

\def\wt{\widetilde}
\usepackage[all]{xy}

\begin{document}

\begin{titlepage}

\begin{flushright}
TU-1225
\end{flushright}

\vskip 3cm

\begin{center}

{\Large \bfseries Constraints on the topology of Type IIB string theory}

\vskip 1cm
Kazuya Yonekura
\vskip 1cm

\begin{tabular}{ll}
Department of Physics, Tohoku University, Sendai 980-8578, Japan
\end{tabular}

\vskip 1cm

\end{center}

\noindent
We discuss some topological constraints on Type~IIB string theory that cannot be described by elementary characteristic classes. Anomaly cancellation of the worldvolume theory of a D3-brane requires a shifted Dirac quantization condition of the Ramond-Ramond 5-form flux. However, the required shift is sometimes inconsistent depending on the topology of spacetime backgrounds.  
The obstruction to the existence of a shifted quantization is described by a degree-6 cohomology class whose definition involves spin structure of spacetime manifolds. The cohomology class is interpreted as a background D3-charge, and the Gauss law constraint requires inclusion of additional D3-branes to cancel it. 

\end{titlepage}

\setcounter{tocdepth}{3}


\tableofcontents

\newpage
\section{Introduction and summary}

Constraints on the topology of string theories are easy to see if they are described by differential forms. However, there are also constraints that are not captured by differential forms.

For example, let us consider the case of heterotic string theories. The field strength 3-form $H_3$ of the Neveu-Schwarz (NS) $B$-field satisfies a modified Bianchi identity of the form\,\footnote{Throughout the paper, all $p$-form fields $G_p$ are normalized in such a way that the ordinary Dirac quantization condition would be given by $\int G_p \in \bZ$ so that there is no factor of $2\pi$. This rule applies also to the $\U(1)$ gauge field on a D-brane. }
\beq
\d H_3 =  c(F) - \lambda (R), \label{eq:introhetbianchi}
\eeq
where $ \lambda (R)$ is given in terms of the Riemann curvature 2-form $R$ as $ \lambda(R) = - \frac14 \tr (R/2\pi)^2$, and $ c(F)$ is given in terms of the gauge field strength 2-form $F$ roughly as $ c(F) \sim \tr F^2$ depending on the details of the theory under consideration. 

Let $[c]_\bR$ and $[\lambda]_{\bR}$ be the de~Rham cohomology classes represented by $c(F)$ and $\lambda(R)$, respectively. Thus they are elements of the real cohomology group $ H^4(X; \bR)$,
where $X$ is the spacetime manifold. The modified Bianchi identity imposes a constraint on the cohomology classes as 
\beq
0=[c]_\bR - [\lambda]_{\bR} \in H^4(X; \bR) . \label{eq:hetdiff}
\eeq
This is the part of constraints that can be described by differential forms.

However, there can be more nontrivial constraints that are not captured by differential forms~\cite{Witten:1985mj}. Assume that the spacetime manifold $X$ is a spin manifold. Then there are integral cohomology classes 
$[c]_\bZ,[\lambda]_\bZ \in H^4(X;\bZ)$ 
whose images in the real cohomology $H^4(X; \bR)$ are given by $[c]_\bR$ and $ [\lambda]_{\bR} $. A constraint that is not visible at the level of differential forms is that the integral cohomology version of \eqref{eq:hetdiff} must be satisfied,
\beq
0=[c]_\bZ - [\lambda]_{\bZ}  \in H^4(X;\bZ) .  \label{eq:hetint}
\eeq
For instance, let us consider a 7-dimensional lens space $S^7/\bZ_k$. Locally this is the same as the sphere $S^7$ and we have (for the standard metric on $S^7$) $ \lambda (R)=0$ as a differential form. On the other hand,  $[\lambda]_{\bZ}$ is one-half of the first Pontryagin class $p_1 \in H^4(X;\bZ)$ and it is nonzero depending on the value of $k$.\footnote{Let $\pi :S^7/\bZ_k \to \mathbb{CP}^3$ be the projection for the fiber bundle $S^1 \to S^7/\bZ_k \to \mathbb{CP}^3$ . The Pontryagin classes on $S^7/\bZ_k$ are obtained by pulling back the corresponding Pontryagin classes on $ \mathbb{CP}^3$ because the tangent direction to the $S^1$-fiber gives a trivial bundle. The total Pontryagin class of $ \mathbb{CP}^3$ is $p( \mathbb{CP}^3) = (1+\sc^2)^4$, where $\sc \in H^2(\mathbb{CP}^3;\bZ)$ is the generator. Hence we have $p_1(S^7/\bZ_k) = 4 (\pi^* \sc)^2$. The cohomology group $H^{2i}(S^7/\bZ_k; \bZ)\simeq \bZ_k$ is generated by $(\pi^* \sc)^i$. Thus $p_1(S^7/\bZ_k) $ is nonzero if $4$ is not divisible by $k$.   } Thus the constraint \eqref{eq:hetint} is nontrivial even when the constraint \eqref{eq:hetdiff} on differential forms is trivial. 

There is a further generalization of \eqref{eq:hetint}. General (possibly nonsupersymmetric) heterotic string theories may be formulated on non-spin manifolds depending on the details of the gauge group. In the most general case, we need a generalized cohomology theory $(I\Omega^{\rm spin})^4(X)$ (whose details are not necessary for the present paper). There are cohomology classes $[c], [\lambda] \in (I\Omega^{\rm spin})^4(X)$ and the constraint is given by~\cite{Yonekura:2022reu}
\beq
0=[c] - [\lambda]  \in (I\Omega^{\rm spin})^4(X).
\eeq 
For instance, if the gauge bundle is trivial and hence $[c]=0$, then the equation $[\lambda]=0$ contains the condition that $X$ admits a spin structure, as well as the condition $[\lambda]_\bZ=0$.\footnote{The condition $[\lambda]=0$ implies that $X$ admits a spin structure. The actual spin structure on $X$ is given by discrete theta terms in the worldsheet theory, and they may be regarded as part of the ``$B$-field'' in the generalized sense~\cite{Yonekura:2022reu}.} For $[c] \neq 0$, $X$ need not be spin nor orientable.

The above constraints for heterotic string theories follow from the anomaly cancellation of the worldsheet theory of a fundamental string~\cite{Witten:1985mj,Witten:1999eg,Yonekura:2022reu}. The differential form condition comes from worldsheet perturbative anomalies, while more nontrivial constraints come from nonperturbative, global anomalies.

String theories contain not only the fundamental string, but also various types of branes. The purpose of the present paper is to study constraints that come from the anomaly of the worldvolume theory of a D3-brane in Type~IIB string theory. Type~IIB string theory has the $\GL(2,\bZ)$ duality symmetry,\footnote{The duality group of Type~IIB string theory is often stated to be $\SL(2,\bZ)$. However, if we add target space symmetries such as $\Omega$ which originates from the worldsheet orientation reversal, the group becomes $\GL(2,\bZ)$. To accommodate actions on fermions, this group must be further enhanced to a $\pin^+$ double cover~\cite{Tachikawa:2018njr}.} and we will consider general $\GL(2,\bZ)$ bundles, or in other words F-theory (but we do not include singular fibers or 7-branes).

One of the main results of the present paper is as follows. Recall that, at the differential form level, the Bianchi identities for the Ramond-Ramond (RR) fields are modified. Let $G_p$ ($p=1,3,5,\cdots$) be the field strength $p$-form of the RR fields, and let $H_3$ be the field strength 3-form of the NSNS $B$-field. We have a modified Bianchi identity for the $G_5$ given by\,\footnote{In this paper, the 3-form $G_3$ is defined such that its Bianchi identity is $\d G_3=0$ rather than $\d G_3= - H_3 \wedge G_1$. This definition is more convenient to make the $\GL(2,\bZ)$ symmetry manifest. }
\beq
\d G_5 = -H_3 \wedge G_3.
\eeq
Let $[H_3 \wedge G_3]_\bR \in H^6(X; \wt \bR)$ be the de~Rham cohomology class of $H_3 \wedge G_3$. (The tilde on $\wt \bR$ will be explained later in the paper.) Then the modified Bianchi identity implies
\beq
0 = [H_3 \wedge G_3]_\bR \in H^6(X; \wt \bR).
\eeq
At the level of integral cohomology, we will obtain a constraint which we write as
\beq
0 = \sC_6 \in  H^6(X; \wt \bZ).
\eeq
The main point of the present paper is to give a definition of the element $\sC_6 \in  H^6(X; \wt \bZ)$.
The image $[\sC_6]_{\bR} \in H^6(X; \wt \bR)$ of $\sC_6$ in the real cohomology is given by $ [H_3 \wedge G_3]_\bR$. The $\sC_6$ contains more information than just the real cohomology class. It will turn out that $\sC_6$ is nontrivial even if other fields such as the RR 2-form $C_2$ and the NSNS 2-form $B_2$ are zero. One of the interesting points is that the definition of $\sC_6$ requires the spin structure of spacetime $X$ or its generalization in the presence of a general $\GL(2,\bZ)$ bundle. In this sense, $\sC_6$ cannot be expressed by elementary characteristic classes such as Pontryagin classes and Stiefel-Whitney classes. 

Along the way of defining $\sC_6$, we need to discuss the concept of shifted quantization condition. The ordinary Dirac quantization condition for $G_5$ is given by $\int_N G_5 \in \bZ$ where $N$ is a closed 5-manifold. However, the quantization condition is modified by the anomaly of the worldvolume theory of a D3-brane as
\beq
\int_N G_5 \in \frac{\i}{2\pi} \log \cZ_{\rm anomaly}(N) + \bZ, \label{eq:introshift}
\eeq
where $ \cZ_{\rm anomaly}(N) \in \U(1)$ is the anomaly of a D3-brane evaluated on $N$, which we explain later in the paper. This kind of modification of the Dirac quantization condition by anomalies is discussed for M2-branes in \cite{Witten:1996md,Witten:2016cio}, and for D-branes in the context of K-theory in \cite{Moore:1999gb}, and in the context of O-plane backgrounds in \cite{Tachikawa:2018njr}. We will discuss more checks of it in the present paper for the case of a D3-brane. The cohomology class $\sC_6$ will be defined as the obstruction to the existence of a shifted quantization \eqref{eq:introshift}. If $\sC_6$ is nonzero, there is no $G_5$ satisfying \eqref{eq:introshift} and hence the anomaly of a D3-brane cannot be cancelled. 

Even if $\sC_6$ is nonzero, we can still consider Type~IIB string theory in that background if we include D3-branes. (These D3-branes are different from the one used to derive \eqref{eq:introshift}.) For example, in the case of heterotic string theories, the presence of NS5-branes modifies the equation \eqref{eq:introhetbianchi} as
\beq
\d H_3 =  c(F) - \lambda (R) + \delta_L,
\eeq
where $L$ is the worldvolume of the NS5-branes, and $\delta_L$ is the delta function 4-form supported on $L$. Thus the condition \eqref{eq:hetdiff} is modified by the Poincare dual of the homology class of $L$. There is a corresponding modification for integral cohomology. In the same way, if we introduce D3-branes on a worldvolume $L \subset X$ such that the homology class of $L$ is the Poincare dual of $\sC_6 \in H^6(X; \wt \bZ)$, we get a consistent background. Thus, $\sC_6$ can be regarded as a background D3-charge. Therefore, whenever we consider backgrounds in which $[c]-[\lambda] \neq 0$ or $\sC_6 \neq 0$, we seriously need to take into account the existence of NS5-branes or D3-branes for heterotic or Type~IIB string theory, respectively.

Our definition of $\sC_6$ uses the anomaly $ \cZ_{\rm anomaly}$, and hence the ability to actually compute it depends on the available knowledge of anomalies. We will give examples in the paper. The understanding of nonperturbative anomalies has been developed both at the conceptual and computational levels. Anomaly inflow~\cite{Callan:1984sa} gives the conceptual framework for general anomalies (see \cite{Freed:2014iua,Monnier:2019ytc} for exposition of general discussions, and \cite{Witten:2015aba,Yonekura:2016wuc,Witten:2019bou} and \cite{Hsieh:2020jpj} for nonperturbative description of anomaly inflow for fermions and $p$-form fields, respectively.) For the classification of anomalies, the concept of bordism turned out to be important \cite{Kapustin:2014tfa,Kapustin:2014dxa,Freed:2016rqq,Yonekura:2018ufj,Yamashita:2021cao} and will also play a role in the present paper. The computations of bordism groups have been done in old mathematics~\cite{StongTextbook} as well as recent physics papers and applied to string theories and particle physics (e.g. \cite{Garcia-Etxebarria:2018ajm,Hsieh:2018ifc,Monnier:2018nfs,BCGuide,Guo:2018vij,Hsieh:2019iba,Wan:2019gqr,Davighi:2019rcd,Cordova:2019wpi,Kaidi:2019tyf,GarciaEtxebarria:2020xsr,Montero:2020icj,Yonekura:2020upo,Lee:2020ojw,Davighi:2020uab,Dierigl:2020lai,Davighi:2020kok,Lee:2020ewl,Wang:2020xyo,Delmastro:2021xox,Kobayashi:2021qfj,Cvetic:2021maf,Grigoletto:2021zyv,Blumenhagen:2021nmi,Lee:2021crt,Wang:2021ayd,Debray:2021vob,Grigoletto:2021oho,Koizumi:2021rpe,Lee:2022spd,Andriot:2022mri,Choi:2022odr,Delmastro:2022pfo,Davighi:2022icj,Blumenhagen:2022bvh,Putrov:2022pua,Debray:2022wcd,Chen:2022cyw,Dierigl:2022zll,Koizumi:2023olk,Davighi:2023luh,Putrov:2023jqi,Basile:2023knk,Debray:2023yrs,Basile:2023zng,Antinucci:2023ezl,Davighi:2023mzg}). In particular, extensive studies of $\GL(2,\bZ)$ duality bundles and bordisms in Type~IIB string theory have been done in \cite{Debray:2021vob,Dierigl:2022reg,Debray:2023yrs,Dierigl:2023jdp} which give one of the important motivations for the present paper.

This paper is organized as follows. The existence of some constraints on Type~IIB string theory can be inferred from the duality between M-theory and Type~IIB string theory, so we review constraints on M-theory in Section~\ref{sec:Mtheory} and give examples in Section~\ref{sec:example1} based on the duality. General relation between anomalies of branes and constraints on topology is discussed in Section~\ref{sec:anomalycohomology}. The application of the general discussions to the case of a D3-brane requires some technical details which we give in Section~\ref{sec:Homsub}. Then we consider the constraints from the anomaly of a D3-brane in Section~\ref{sec:IIBconstraint}, where the cohomology class $\sC_6$ is defined. More examples, some of which are essential for the definition of $\sC_6$, will be discussed in Section~\ref{sec:ex2}.

\section{Constraints on M-theory}\label{sec:Mtheory}

The existence of some constraints on Type~IIB string theory or F-theory (with general $\GL(2,\bZ)$ duality bundles but without singular fibers or 7-branes) can be inferred from its relation to M-theory. In Section~\ref{sec:example1}, we will construct an explicit example by using this relation. Therefore, we review constraints on M-theory~\cite{Witten:1996md,Witten:2016cio,Freed:2019sco}. 

\subsection{Shifted quantization}\label{sec:Mtheoryshift}
M-theory can be formulated on manifolds that are not necessarily orientable. (More precisely, it can be formulated on manifolds with $\pin^{+}$ structure.) On a spacetime manifold $X$ which is possibly non-orientable, let $\zeta$ be the $\bZ_2$ bundle associated to the determinant ${\rm Det}\, TX$ of the tangent bundle $TX$; the transition function of $\zeta$ is $\pm 1$ depending on whether the corresponding transition function of $TX$ changes orientation or not. We call $\zeta$ the orientation bundle of $X$ in this section. (In Type~IIB string theory, $\zeta$ will have a slightly different meaning.) By using it, we also define bundles
\beq
\wt \bZ = \zeta \times_{\bZ_2} \bZ, \qquad \wt \bR = \zeta \times_{\bZ_2}  \bR , \qquad \wt {\bR/\bZ} = \zeta \times_{\bZ_2}  (\bR/\bZ) \label{eq:twistZ}
\eeq
whose fibers are $\bZ$, $\bR$, and $\bR/\bZ$ respectively, and transition functions $\pm 1$ are the same as $\zeta$. Then we can consider cohomology groups $H^i(X; \wt \bZ)$, $H^i(X; \wt \bR)$ and $H^i(X; \wt {\bR/\bZ})$ whose coefficients are twisted by $\zeta$.\footnote{For some background knowledge of algebraic topology required throughout the paper, see e.g. \cite{Hatcher1,MilnorStasheff:MR0440554}.}

M-theory contains a 3-form field $C_3$ whose field strength $G_4$ is a 4-form. It changes sign under orientation reversal, so the naive Dirac quantization condition is that the de~Rham cohomology class $[G_4]$ of $G_4$ takes values in (the image in the de~Rham cohomology of) $H^4(X; \wt \bZ)$. However, this is not the case. The quantization condition for $G_4$ is known to be shifted as follows. Let $w_i \in H^i(X; \bZ_2)$ ($i=1,2,\cdots$) be the  Stiefel-Whitney classes of the tangent bundle $TX$. Then, roughly, $G_4$ is quantized as
\beq
[G_4] \in \frac{1}{2} w_4 + H^4(X; \wt \bZ). \label{eq:Mshifted}
\eeq
We will explain the more precise meaning of this equation in Section~\ref{sec:Mobs}.

Let us sketch one of the reasons for the condition \eqref{eq:Mshifted}. (See \cite{Witten:2016cio} for the details of what follows.) It follows from the anomaly cancellation condition of the worldvolume theory of an M2-brane. 

An M2-brane worldvolume $M$ contains fermions. The Lie algebra of the structure group of the normal bundle $\nu(M)$ to $M$ is $\so(8)$, and the fermions transform in one of the irreducible spinor representations of $\so(8)$. Also, the worldvolume theory has parity or orientation reversal symmetry, and hence the worldvolume $M$ is possibly non-orientable. There is a parity anomaly~\cite{Redlich:1983dv,Niemi:1983rq,Alvarez-Gaume:1984zst} between the $\so(8)$ symmetry and the orientation reversal symmetry, as well as a purely gravitational anomaly for the orientation reversal symmetry~\cite{Witten:2015aba}. 

In general, anomalies can be described by anomaly inflow~\cite{Callan:1984sa} as follows. Anomalies are ambiguities of the phase of partition functions. In the case of an M2-brane, suppose we take a submanifold $N' \subset X$ such that its boundary is given by the worldvolume $M$ of the M2-brane, 
\beq
\partial N'  = M.
\eeq 
Then the partition function of the worldvolume theory can be defined without any phase ambiguity by introducing an appropriate term on $N'$.\footnote{See \cite{Witten:2019bou} for detailed discussions of the nonperturbative anomaly inflow mechanism for fermions.} 
However, it depends on a choice of $N' $ even though the worldvolume theory is living on $M$. Let us denote the partition function of the worldvolume theory defined by using $N'$ as $\cZ_\textrm{matter}(N' )$. We can consider another manifold $N''$ and the corresponding partition function $\cZ_\textrm{matter}(N'')$. In the case of a single M2-brane, their ratio turns out to be given by
\beq
\frac{\cZ_\textrm{matter}(N'')}{\cZ_\textrm{matter}(N')} = \exp\left( \pi \i \int_{N} w_4 \right), \qquad N = N'' \cup \overline{N'},
\eeq
where $ N \subset X$ is a closed submanifold constructed by gluing $N'$ and $N''$ along their common boundary $M$.\footnote{The $\overline{N'}$ is the reversal of the structure on $N'$. See \cite{Freed:2016rqq,Yonekura:2018ufj} for explanation of this concept.}

We want the partition function of the worldvolume theory to depend only on $M$ rather than its artificial extension $N'$. This is achieved as follows. The M2-brane worldvolume has a coupling to the 3-form field $C_3$. It is roughly given by $\exp( 2\pi \i \int_M C_3)$. More precisely, by taking $N'$ with $\partial N' =M$, we define this coupling as $\exp( 2\pi \i \int_{N'} G_4)$ so that the total partition function is
\beq
\cZ(N') : = \cZ_\textrm{matter}(N') \exp\left( 2\pi \i \int_{N'} G_4 \right).
\eeq
Now we can ask whether $\cZ(N')$ depends on a choice of $N'$ or not. By taking another $N''$, we consider the ratio as above and get
\beq
\frac{\cZ(N'')}{\cZ(N')} = \exp\left( 2\pi \i \int_{N} \left(\frac12 w_4+G_4 \right) \right), \qquad N = N'' \cup \overline{N'}.
\eeq
If \eqref{eq:Mshifted} is satisfied, we obtain $\cZ(N'')=\cZ(N')$ and hence it does not depend on a choice of $N'$. 

\subsection{Obstruction to shifted quantization}\label{sec:Mobs}
In some cases, there is an obstruction to the existence of a $G_4$ satisfying the shifted quantization condition \eqref{eq:Mshifted}. Before explaining it, however, we need some preparation about cohomology and homology. 

For the local coefficient systems \eqref{eq:twistZ}, we have a short exact sequence
\beq
0 \to \wt \bZ \to \wt \bR \to  \wt {\bR/\bZ}  \to 0
\eeq
which gives the long exact sequence of cohomology groups
\beq
\cdots \to H^n(X; \wt \bZ ) \to  H^n(X; \wt \bR ) \xrightarrow{\wt \alpha} H^n(X; \wt  {\bR/\bZ}  ) \xrightarrow{\wt \beta} H^{n+1}(X;\wt \bZ ) \to \cdots \label{eq:longexact}
\eeq
where 
\beq
\wt \beta : H^n(X; \wt  {\bR/\bZ}  ) \to H^{n+1}(X; \wt \bZ ). \label{eq:bockstein}
\eeq
is the Bockstein homomorphism, and 
\beq
\wt \alpha : H^n(X; \wt \bR ) \to H^n(X; \wt  {\bR/\bZ}  ) .\label{eq:alphamap}
\eeq
 is defined in the obvious way from $\wt \bR \to  \wt {\bR/\bZ}$.

Another fact we need is that a 4-manifold $N$ that appeared in the above discussion represents a homology element which we denote as $[N]$,
\beq
[N] \in H_4(X; \wt \bZ). \label{eq:homologyN}
\eeq
The coefficient is understood as follows. The normal bundle $\nu(M)$ to the worldvolume $M$ must be oriented for the M2-brane worldvolume theory to make sense because the fermions are in one of the irreducible spinor representations of $\so(8)$ (i.e. they have a definite ``chirality'' of the normal bundle). For the anomaly inflow to be possible, the $N$ must have the same structure. Thus the normal bundle $\nu(N)$ must be oriented, and the orientation bundle of $X$ restricted to $N$ is the same as the orientation bundle for $N$. This is the reason for the twisted coefficient in \eqref{eq:homologyN}.

By using the identification $\U(1) = \bR/\bZ$, we regard
$ \exp\left( \pi \i \int_{N} w_4 \right) \in \U(1)$ as $\frac12  \int_{N} w_4 \in \bR/\bZ$. The map
\beq
\wt w'_4 :  H_4(X; \wt \bZ) \ni [N] \mapsto \frac12  \int_{N} w_4 \in \bR/\bZ
\eeq
gives an element of
\beq
\Hom( H_4(X; \wt \bZ) , \bR/\bZ) = H^4(X; \wt {\bR/\bZ}).
\eeq
In this way we obtain an element of the cohomology group
\beq
\wt w'_4 \in H^4(X, \wt {\bR/\bZ}).
\eeq
In the same way, from the map
\beq
[G_4] : H_4(X; \wt \bZ) \ni [N] \mapsto \int_{N} G_4 \in \bR,
\eeq
we get an element
\beq
[G_4] \in H^4(X; \wt \bR).
\eeq
This is of course the de~Rham cohomology class of $G_4$.

Now we can understand the precise meaning of the shifted quantization condition \eqref{eq:Mshifted}. From the discussion of the anomaly cancellation, it should be interpreted as the statement that
\beq
\wt \alpha ([G_4]) = \wt w'_4, \label{eq:shifttedincohomology}
\eeq
where $\wt \alpha$ is the map \eqref{eq:alphamap}. 

Let us define $ \sC_5$ by
\beq
\sC_5 = \wt \beta ( \wt w'_4) \in H^5(X; \wt \bZ). \label{eq:defW5}
\eeq
This is a characteristic class which is determined completely by the properties of $X$, i.e., it does not depend on $G_4$. For some manifolds, $\sC_5$ is nonzero and we will give an example in Section~\ref{sec:example1}. 

For the shifted quantization condition to be possible (without inclusion of M5-branes as we discuss shortly), the spacetime manifold $X$ must be such that $\sC_5=0$. This is because if \eqref{eq:shifttedincohomology} is satisfied, then the exact sequence \eqref{eq:longexact} implies that
\beq
\sC_5 = \wt \beta ( \wt w'_4)  = \wt \beta (\wt \alpha ([G_4]))  =0.
\eeq
In other words, $\sC_5$ is the obstruction to the existence of a shifted quantization of $G_4$. 

\subsection{Inclusion of branes}\label{sec:Minc}

Is a manifold $X$ useless for M-theory if $\sC_5 \neq 0$? We can make sense M-theory on $X$ if we introduce M5-branes. The Bockstein map $\wt \beta$ is defined by using the coboundary map, and $\wt \beta  \wt \alpha$ may be roughly regarded as the ``exterior derivative'' acting on $G_4$. Then, roughly, we write
\beq
\d G_4 = \sC_5.
\eeq
This equation implies that $ \sC_5$ is an ``exact form'' and hence its cohomology class is zero. However, suppose that we have an M5-brane whose worldvolume is $L$. Let $\delta_L $ be the delta function 5-form supported on $L$. (More explicitly, if the submanifold $L$ is described in terms of local coordinates $x^0,x^1 ,\cdots x^{10}$ as $x^6=\cdots =x^{10}=0$, then $\delta_L = \pm \delta(x^6) \d x^6 \wedge \cdots \wedge \delta(x^{10}) \d x^{10} $ where the sign is determined by the orientation of the M5-brane.) Then the above equation is modified to
\beq
\d G_4 = \sC_5 + \delta_L.
\eeq
Therefore, if the homology class
\beq
[L] \in H_6(X; \bZ)
\eeq
is (the negative of) the Poincare dual of $ \sC_5 \in H^5(X; \wt \bZ)$, the equation for $G_4$ can be solved without a topological obstruction. 

The above rough discussions may be made more precise by using differential cohomology theory~\cite{CS,Hopkins:2002rd} (see  \cite{Freed:2006yc,Hsieh:2020jpj} for reviews aimed at physicists). We regard $\sC_5$ and $\delta_L$ as (part of) differential cocycles, $G_4$ as a differential cochain, and $\d$ as the differential coboundary map. We omit the details.

\section{Examples in M-theory and Type~IIB string theory}\label{sec:example1}

In this section, we construct examples in Type~IIB string theory such that (i) the quantization condition for the Ramond-Ramond (RR) 5-form field strength $G_5$ is shifted, and (ii) the required shift is inconsistent, or in other words there is an obstruction to the shifted quantization condition. These examples are constructed by starting from M-theory and then using the duality to Type~IIB string theory or F-theory.\footnote{For reviews of F-theory, one may consult e.g. \cite{Denef:2008wq,Heckman:2010bq,Taylor:2011wt,Weigand:2018rez}. } 

Before giving examples, let us make a remark about $G_5$. In Lorentzian signature manifolds, $G_5$ satisfies the self-dual condition $G_5 = \star G_5$ where $\star$ is the Hodge star. This condition is very subtle in Euclidean signature spacetimes~\cite{Witten:1996hc} (see \cite{Diaconescu:2000wy} for a very detailed study for the case of Type~IIA string theory). However, when the spacetime $X$ is of the form $X = \bR \times X'$ or $S^1 \times X'$, we can just consider the components of the 5-form whose indices are all in the direction $X'$. Then the other components are determined by the self-duality condition. Our examples below are of this type. Throughout the present paper, we do not seriously take into account the self-duality condition, and  postpone it for future work. 

\subsection{Shifted quantization}\label{sec:exshift}
We consider the quaternionic projective space $\mathbb{ HP}^n$.  (See \cite{MilnorStasheff:MR0440554} for some of the relevant facts used below.) In the following discussions, it is convenient to regard a quaternion number $q \in \bH$ to be a $2 \times 2$ complex matrix of the form $q = a -\i b \sigma_1-\i c \sigma_2 -\i c \sigma_3$, where $a,b,c,d$ are real numbers and $\sigma_{1,2,3}$ are Pauli matrices. The absolute value of $q$ is given by $|q|^2=a^2+b^2+c^2+d^2$.

$\mathbb{ HP}^n$ is defined as the set of nonzero vectors $\vec q  =(q_1, \cdots, q_{n+1}) \in \bH^{n+1}$, with the equivalence relation $(q_1, \cdots, q_{n+1}) \sim (q_1 g, \cdots, q_{n+1}g)$ for nonzero $g \in \bH$. The equivalence classes are denoted as 
\beq
[q_1, \cdots, q_{n+1}] \in \mathbb{HP}^{n}.
\eeq
It is possible to assume that $|\vec q\,|^2=\sum_{i=1}^{n+1}|q_i|^2=1$ and $|g|=1$. A quaternion $g$ with $|g|=1$ is an element of $\SU(2)$. Thus there is a structure of a principal fiber bundle $\SU(2) \to S^{4n+3} \to \mathbb{HP}^{n}$ whose fiber is $\SU(2)$, the base is $\mathbb{HP}^{n}$, and the total space is $S^{4n+3}$. We call this the canonical $\SU(2)$ bundle.

Let $\sc$ be the second Chern class of the canonical $\SU(2)$ bundle. It is known that the cohomology ring of $\mathbb{HP}^{n}$ is generated by $\sc$, meaning that $H^{4i}( \mathbb{HP}^{n};\bZ) \simeq \bZ$ ($i=0,1,\cdots,n$) is generated by $\sc^i$, and other $H^j( \mathbb{HP}^{n}; \bZ)$ ($j \notin 4 \bZ$) are zero. For our purposes, it is enough to notice the following. $\mathbb{HP}^{n}$ has a submanifold $\mathbb{HP}^{1}$ of the form $[q_1, q_2, 0,\cdots, 0]$. The $\mathbb{HP}^{1}$ is just $S^4$ since it is constructed from two disks 
$D^4 =\{[q,1]]~|~q \in \bH, ~ |q|^2 \leq 1\} $ and $D'^4 = \{[1, q']~|~q' \in \bH, ~|q'|^2 \leq 1\}$
by gluing them along the boundaries $|q|=|q'|=1$ by $q'=q^{-1}$. The transition function $g$ in the gluing is determined by $(1, q' )g = (q,1)$ and hence we have $g=q=q'^{-1}$. This is precisely the transition function of the $\SU(2)$ bundle with instanton number 1 on $S^4$, because it gives the generator of $\pi_3(\SU(2)) \simeq \bZ$. Thus we have $\int_{S^4} \sc =1$ (after changing orientation if necessary). 

In general, suppose we have a complex vector bundle $\cE_\bC$, and let $\cE_\bR$ be the underlying real vector bundle. Then the Stiefel-Whitney classes $w_i(\cE_\bR)$ of $\cE_\bR$ are related to the Chern classes $c_i(\cE_\bC)$ of $\cE_\bC$ by
\beq
w_{2i}(\cE_\bR) = c_i(\cE_\bC) \mod 2, \qquad w_{2i-1}=0. \label{eq:CWS}
\eeq
In particular, for the canonical $\SU(2)$ bundle on $\mathbb{HP}^{n}$, let $\cE_{\bC}$ and $\cE_{\bR}$ be the associated complex rank 2 bundle and its underlying real rank 4 bundle. The fourth Stiefel-Whitney class $w_4(\cE_\bR)$ is given by the second Chern class $\sc = c_2(\cE_\bC)$ modulo 2.

For our purposes we use $\mathbb{HP}^2$. Consider the submanifold 
\beq
S^4=\mathbb{HP}^1 \subset \mathbb{HP}^2.
\eeq
On this submanifold, the tangent bundle $T  \mathbb{HP}^2$ splits into the tangent bundle $T S^4$ of $S^4$ and the normal bundle $\nu(S^4)$.

The normal bundle is precisely the rank 4 bundle $\cE_\bR$ associated to the canonical $\SU(2)$ bundle. The neighborhood of $S^4=\mathbb{HP}^1$ is described by $[q_1, q_2, q_3]$ with $|q_3| \ll 1$ and hence $q_3$ is the normal direction, and the transition function for $q_3$ is determined by $(1,q' ,q'_3) g=(q,1,q_3)$ where $g=q =q'^{-1}$ as above. The spin bundle associated to the normal bundle is such that under $\so(4) = \su(2) \times \su(2)$, one $\su(2)$ has instanton number 1 and the other has instanton number 0.\footnote{Recall that $(g_1, g_2) \in \SU(2) \times \SU(2)$ acts on $q_3 \in \bH$ as $ q_3 \to g_1 q_3 g_2^{-1}$.}

The tangent bundle $TS^4$ of $S^4$ becomes trivial if we add a trivial bundle of rank 1, and its Stiefel-Whitney classes are zero. Thus, on $S^4$, the Stiefel-Whitney class $w_4(\mathbb{HP}^2)$ is given by
\beq
w_4(\mathbb{HP}^2)  = w_4(\cE_\bR) =\sc \mod 2 
\eeq
and in particular
\beq
\int_{S^4} w_4(\mathbb{HP}^2)  =1 \in \bZ_2. \label{eq:HP2w4}
\eeq

Now we consider M-theory on the 11-manifold $\bR \times T^2 \times \mathbb{HP}^2$ where on the torus $T^2$ we take periodic spin structures in both directions. The value \eqref{eq:HP2w4} is nontrivial and hence we have a shifted quantization condition of $G_4$,
\beq
\int_{S^4} G_4 \in \frac12 + \bZ. \label{eq:HP2shift}
\eeq
M-theory on $\bR \times T^2 \times \mathbb{HP}^2$ is dual to Type~IIB string theory on $\bR \times S^1 \times \mathbb{HP}^2$ where $S^1$ has the periodic spin structure.\footnote{If we forget the duality to M-theory, we can consider Type~IIB string theory on $\bR \times S^1 \times \mathbb{HP}^2$ with the anti-periodic spin structure on $S^1$. In this case, the shift \eqref{eq:HP2shiftIIB} does not appear as we discuss in Section~\ref{sec:global}. } The integral $\int_{S^4}G_4$ in M-theory is dual to the integral $\int_{S^1 \times S^4} G_5$, where $G_5$ is the 5-form field strength in Type~IIB string theory. Therefore, we get
\beq
\int_{S^1 \times S^4} G_5  \in \frac12 + \bZ. \label{eq:HP2shiftIIB}
\eeq
Thus the quantization condition for $G_5$ is shifted.

\subsection{Obstruction to shifted quantization}\label{sec:exobs}

Let ${\rm KB}$ be the Klein bottle, and $K$ be a 3-manifold given by $K = {\rm KB} \times S^1$.
As our next example, we consider M-theory on  $K \times \mathbb{HP}^2$.

It is convenient to regard $K$ as a fiber bundle
$T^2 \to K \to S^1$ whose fiber is $T^2$ and the base is $S^1$. To distinguish several $S^1$'s, we denote $T^2=S^1_A  \times S^1_B$ for the fiber and $\widetilde S^1$ for the base. The fiber bundle structures for ${\rm KB}$ and $K$ are given by
\beq
  \xymatrix{
    S^1_A \ar[r] & {\rm KB} \ar[d] \\
    & \widetilde S^1
  }, \qquad
  \xymatrix{
    T^2 = S^1_A \times S^1_B \ar[r] & K = {\rm KB} \times S_B  \ar[d] \\
    & \widetilde S^1
  }, \label{eq:fiberstr}
\eeq
Thus $S^1_A$ is nontrivially fibered over $\wt S^1$, while $S^1_B$ is trivially fibered. 

Let $p$ be an arbitrary point on $K$. On $ \{p\} \times S^4 \subset K \times \mathbb{HP}^2$, we have the condition \eqref{eq:HP2shift}. However, now we encounter a problem. Let us gradually change the position of $p \in K$ and let it go around $\wt S^1$ once. We denote the coordinate of $\wt S^1$ as $\theta$ which has periodicity $2\pi$. When we go from $\theta=0$ to $\theta =2\pi$, the orientation of $K$ is flipped (i.e. the orientation bundle has a nontrivial holonomy around $\wt S^1$). Because $G_4$ is odd under orientation reversal, we have 
\beq
G_4|_{\theta =2\pi} = -G_4|_{\theta = 0} .  \label{eq:kuru}
\eeq
However, in a continuous change of $\theta$, the quantized value of the integral \eqref{eq:HP2shift} cannot change, so 
\beq
\int_{ S^4}  G_4|_{\theta}=\frac{1}{2} +n \label{eq:HP2shift2}
\eeq
for any $\theta$, where $n \in \bZ$ is an integer. These two equations require
\beq
\left(\frac{1}{2} +n \right) = -\left(\frac{1}{2} +n \right).
\eeq
This is inconsistent.

M-theory on $K \times \mathbb{HP}^2$ is dual to Type~IIB string theory on $S_C^1 \times \wt S^1 \times \mathbb{HP}^2$. (More precisely, we take the spin structure on $K$ such that this duality holds.) We have a new circle $S_C^1$. The information of the nontrivial fiber structure \eqref{eq:fiberstr} of $T^2$ is encoded in the Type~IIB side as follows. In M-theory, we have $\GL(2,\bZ)$ symmetry acting on $T^2$. This corresponds to the Type~IIB $\GL(2,\bZ)$ duality symmetry. The above $T^2$ fiber bundle structure corresponds to a nontrivial holonomy
\beq
\begin{pmatrix} -1 & 0 \\ 0 & 1 \end{pmatrix} \in \GL(2,\bZ)
\eeq
around $\widetilde S^1$. This corresponds (up to conjugation) to a nontrivial holonomy of the target space $\bZ_2$ symmetry $\Omega$ that comes from the worldsheet orientation reversal symmetry. (We can also use another $\bZ_2$ symmetry $(-1)^{\sF_L}$ since $\Omega$ and $(-1)^{\sF_L}$ are equivalent under conjugation by $S$-duality.) The 5-form field strength $G_5$ is odd under $\Omega$ (and $(-1)^{\sF_L}$). 

The condition \eqref{eq:HP2shiftIIB} in the current case involves $S^1_C$ as
\beq
\int_{S^1_C \times S^4} G_5  \in \frac12 + \bZ.
\eeq
By the same argument as in M-theory, it is inconsistent with the $\Omega$ holonomy around $\wt S^1$.

\subsection{Inclusion of branes}

Let us return to M-theory on $K \times \mathbb{HP}^2$, and now consider inclusion of M5-branes as follows. We take another sphere $S'^4 \subset \mathbb{HP}^2$ such that the intersection number between $S^4$ and $S'^4$ is 1. For instance, these two spheres may be taken to be the subspaces whose elements are of the form $[q_1,q_2,0]$ and $[q_1, 0, q_3]$, respectively. Then they intersect at a single point $[1,0,0]$. Actually, $S^4$ and $S'^4$ are homotopic and the following discussion is valid even if we take $S'^4=S^4$, but it may be more intuitive to distinguish them. 

We put $n$ M5-branes on
\beq
T^2 \times \{\theta_0\} \times S'^4,\label{eq:M5volume}
\eeq
where $\theta_0 \in \wt S^1$ is a point such that $0< \theta_0 < 2\pi$. These M5-branes create a flux of $G_4$. Let $\epsilon>0$ be a small number. Then the flux created by the $n$ M5-branes has the effect that
\beq
\int_{ S^4}  G_4|_{\theta_0 + \epsilon} -  \int_{ S^4}  G_4|_{\theta_0 - \epsilon} = n .
\eeq
This is due to the fact that $[\theta_0-\epsilon, \theta_0+\epsilon] \times S^4$ and the worldvolume \eqref{eq:M5volume} intersect at a single point.
We have $\int_{ S^4}  G_4|_{\theta_0 - \epsilon} =\int_{ S^4}  G_4|_{\theta=0} $ and $\int_{ S^4}  G_4|_{\theta_0 + \epsilon} =\int_{ S^4}  G_4|_{\theta=2\pi} $, and hence the condition \eqref{eq:kuru} requires that
\beq
\int_{ S^4}  G_4|_{\theta=0}  = - \frac{n}{2}.
\eeq
This is consistent with the shifted quantization \eqref{eq:HP2shift} if $n$ is odd.

We remark that the number $n$ of M5-branes is conserved only modulo 2 by the following reason. Let us take one of the $n$ M5-branes, and let it go once around $\wt S^1$. Because of the orientation flip, the M5-brane comes back as an anti-M5-brane. Then it can be pair-annihilated with another M5-brane and the remaining number of M5-branes is $n-2$. Thus, only $n ~{\rm mod}~2$ is topologically invariant. 

Let us compute the $\sC_5$ defined in \eqref{eq:defW5} and check that it is Poincare dual to the M5-brane worldvolume. Essentially it is determined by a simple computation on $\wt S^1$. Consider the exact sequence
\beq
\cdots \to H^0(\wt S^1; \wt \bR) \to H^0(\wt S^1; \wt {\bR/\bZ}) \xrightarrow{\wt \beta} H^1(\wt S^1; \wt \bZ) \to \cdots \label{eq:int1}
\eeq
where the coefficients $\wt \bZ, \wt \bR$ and $\wt {\bR/\bZ}$ are twisted by a bundle which has a nontrivial holonomy around $\wt S^1$. We will also use
\beq
H^i(\wt S^1; \wt \bR) = \Hom( H_i(\wt S^1; \wt \bZ), \bR), \qquad  H^i(\wt S^1; \wt {\bR/\bZ} ) = \Hom( H_i(\wt S^1; \wt \bZ), \bR/\bZ).\label{eq:int2}
\eeq
Let us consider a one-dimensional chain $\sigma_1 : [0,1] \to \wt S^1$ given by $\sigma_1(t) = 2\pi t$ ($t \in [0,1]$). Its boundary is given by
$
\partial \sigma_1 = \sigma_{0, \theta=2\pi} -\sigma_{0, \theta=0} 
$
where $\sigma_{0,\theta}$ is a zero-dimensional chain at the point $\theta \in \wt S^1$. The twisted coefficient $\wt \bZ$ gives
$
\sigma_{0, \theta=2\pi} = - \sigma_{0, \theta=0}
$.
Thus we see that $\partial \sigma_1 = -2\sigma_{0, \theta=0} $, and from this fact we conclude that
\beq
H_0(\wt S^1; \wt \bZ) \simeq \bZ_2.
\eeq
Compare this discussion to the above discussion that the number $n$ of M5-branes is conserved only modulo 2. 
Now \eqref{eq:int2} implies
\beq
H^0(\wt S^1; \wt \bR) \simeq 0, \qquad H^0(\wt S^1; \wt {\bR/\bZ} ) \simeq \bZ_2.
\eeq
Therefore, from \eqref{eq:int1} we get  
\beq
0\to \bZ_2  \xrightarrow{\wt \beta} H^1(\wt S^1; \wt \bZ) 
\eeq
This implies that $\wt \beta(1) \neq 0$.

For $K$ which is the fiber bundle \eqref{eq:fiberstr}, we still have $H_0(K; \wt \bZ) \simeq \bZ_2$ and get
\beq
0\to H^0(K; \wt {\bZ/\bR}) \simeq \bZ_2  \xrightarrow{\wt \beta} H^1(K; \wt \bZ) .
\eeq
For $K \times \mathbb{HP}^2$, the K\"unneth formula and the cohomology groups of $ \mathbb{HP}^2$ mentioned in Section~\ref{sec:exshift} and the universal coefficient theorem imply that for $\bA = \bZ, \bR$ and $\bR/\bZ$ we have
\beq
H^i(K \times \mathbb{HP}^2; \wt \bA) &= \bigoplus_{j=0}^i H^{j}( \mathbb{HP}^2; H^{i-j}( K; \wt \bA))  \nonumber \\
&= H^i(K ; \wt \bA) \oplus H^{i-4}(K ; \wt \bA) \oplus H^{i-8}(K ; \wt \bA),
\eeq
where $H^j$ is defined to be zero for $j<0$. Thus, we get
\beq
0\to H^4(K \times \mathbb{HP}^2; \wt {\bZ/\bR}) \simeq \bZ_2  \xrightarrow{\wt \beta} H^5(K \times \mathbb{HP}^2; \wt \bZ) .
\eeq
Notice that $\wt w'_4 \in H^4(K \times \mathbb{HP}^2; \wt {\bZ/\bR}) $ is a nontrivial element because of \eqref{eq:HP2w4}. Therefore, $\sC_5 = \wt \beta  (\wt w'_4)$ is nonzero. 

Poincare duality implies that $H^1(\wt S^1; \wt \bZ) = H_0(\wt S^1; \wt \bZ) \simeq \bZ_2$. Let $ \sa \in H^1(\wt S^1; \wt \bZ) $ be the unique nonzero element of $H^1(\wt S^1; \wt \bZ)$. Since $\sa$ is unique, we have $\wt \beta(1) = \sa$ on $\wt S^1$. By the projection map $K \to \wt S^1$ of the fiber bundle, we can pullback $\sa$ to $K$ and we denote it by the same symbol $\sa$. Then, from the above structure of the cohomology groups, we see that $\sC_5$ on $K \times \mathbb{HP}^2$ is given by
\beq
 \sC_5 = \sa \sc
\eeq
where $\sc$ is (the pullback of) the generator $\sc \in H^4( \mathbb{HP}^2; \bZ) $. The Poincare dual of $\sC_5 = \sa \sc$ is precisely the cycle \eqref{eq:M5volume} which was used as the worldvolume of M5-branes.

In Type~IIB string theory on $S^1_C \times \wt S^1 \times  \mathbb{HP}^2$, the M5-branes on \eqref{eq:M5volume} are dual to D3-branes on
\beq
\{\phi_0\} \times  \{\theta_0\} \times S'^4 \subset S^1_C \times \wt S^1 \times  \mathbb{HP}^2, \label{eq:D3in}
\eeq
where $\phi_0 \in S^1_C$ is a point on $S^1_C$.

As an aside, let us comment on an important role of the D3-branes wrapped on \eqref{eq:D3in} in the cancellation of an anomaly of the 10-dimensional fields. There are gravitino fields in Type~IIB string theory, and they have a single zero mode in the background $S_C^1 \times \wt S^1 \times \mathbb{HP}^2$~\cite{Lee:2022spd}. As a result, the path integral measure is not invariant under the fermion parity symmetry $(-1)^{\sF}$ and hence a gravitino anomaly exists. On the other hand, we have seen above that we need odd number of D3-branes wrapped on \eqref{eq:D3in}. As we discussed in Section~\ref{sec:exshift}, the spin bundle associated to the normal bundle to $S^4 = \mathbb{HP}^1$ in $\mathbb{HP}^2$ has a single instanton number. The fermions on the worldvolume of the D3-branes are coupled to the spin bundle associated to the normal bundle, and hence the number of zero modes on the $n$-coincident D3-branes is odd (regardless of the worldvolume $\U(n)$ gauge field) if $n$ is odd. The D3-brane path integral measure is not invariant under $(-1)^{\sF}$. This precisely cancels the anomaly of the gravitino mentioned above. 

Anomalies of the 10-dimensional fields in Type~IIB string theory have been studied in \cite{Debray:2021vob}. We leave it for future work to study more systematically the effects of the constraints of the present paper to the anomalies of the 10-dimensional fields.

\subsection{The case that can be related to M-theory}
By using the duality between M-theory and F-theory, we can obtain the following constraints. We consider F-theory with a general $\GL(2,\bZ)$ duality bundle, but without singular fibers (i.e. 7-branes). Suppose that the base 10-dimensional spacetime is of the form $X_{10} = S^1 \times X_{9}$ such that the $S^1$ has the periodic spin structure, and the $\GL(2,\bZ)$ holonomy around $S^1$ is trivial. This is dual to M-theory on $X_{11}$ that is a fiber bundle $T^2 \to X_{11} \to X_{9}$ whose fiber is $T^2$ and the base is $X_9$. The $T^2$ fibration is determined by the $\GL(2,\bZ)$ bundle on $X_9$. 

Let $\zeta$ be the $\bZ_2$ bundle associated to the $\GL(2,\bZ)$ bundle obtained by taking the determinant in $\GL(2,\bZ)$. We can define twisted coefficients of cohomology as in \eqref{eq:twistZ} by using $\zeta$. 

Let $\cV$ be the rank-2 real vector bundle associated to the $\GL(2,\bZ)$ bundle in the defining representation of $\GL(2,\bZ)$. We can define a cohomology class
\beq
\sC_5(\cV \oplus TX_9) \in H^5( X_9; \wt \bZ)
\eeq
for the real bundle $\cV \oplus TX_9$ in the same way as before. The constraint on F-theory in the case of spacetime $S^1 \times X_{9}$ is that $\sC_5(\cV \oplus TX_9)$ must vanish, or we must include D3-branes to cancel it.

Although we could construct explicit examples by using the duality between M-theory and F-theory, the description in terms of $\sC_5(\cV \oplus TX_9) $ is not a final answer for at least two reasons. 

First of all, we can use the relation to M-theory only for Type~IIB backgrounds of the form $S^1 \times X_9$. It is not obvious how to generalize the results to more general backgrounds.

Even in the case $S^1 \times X_9$, there is a point that suggests that the generalization is not straightforward. The relation between the M-theory 4-form $G_4$ and the Type~IIB string theory 5-form $G_5$ is given roughly by $G_4 = \int_{S^1} G_5$. This implies that a shifted quantization condition for $G_5$ and an obstruction to the existence of shifted quantization must involve the $S^1$ direction. However, the tangent bundle $TS^1$ of $S^1$ is topologically trivial, and hence all characteristic classes of $T(S^1 \times X_9) = TS^1 \oplus TX_9$ and the $\GL(2,\bZ)$ duality bundle are the pullback of characteristic classes on $X_9$. They become zero when integrated over $S^1$. Therefore, we expect that the conditions cannot be described by simple characteristic classes. Actually, it will turn out that spin structure is important.

In the case of M-theory, one of the ways to see the shifted quantization condition is to consider the anomaly of the worldvolume theory of an M2-brane. Thus, in the case of Type~IIB string theory, we expect that the anomaly of the worldvolume theory of a D3-brane plays an important role. This is the strategy we take in the present paper.

\section{Anomalies and cohomology}\label{sec:anomalycohomology}

The discussions of the shifted quantization condition from the anomaly of an M2-brane as reviewed in Section~\ref{sec:Mtheoryshift} may be generalized to more general branes. However, let us make remarks about some subtleties. 
\begin{enumerate}

\item 
We want to determine a shifted quantization condition for $\int_{C} G_p$, where $G_p$ is the field strength $p$-form which we are interested in, and $C \in H_p(X ; \wt \bZ)$ is an element of a homology group (whose coefficient is possibly twisted depending on what brane we consider). Then we need to know whether any $C$ can be realized by a map $f: N \to X$ as $C=f_* [N]$ where $N$ is a manifold on which we compute anomalies, and $[N] $ is the fundamental homology class of $N$. 

\item
A worldvolume $M$ of a brane in a spacetime manifold $X$ must be a submanifold if we want to use simple low energy effective field theory on it. In other words, the map $M \to X$ must be an embedding.\footnote{An exception is the fundamental string, for which any smooth map $M \to X$ can be described by the worldsheet theory. In this case the subtleties discussed here are absent. See \cite{Yonekura:2022reu} for the case of general heterotic string theories.} We compute anomalies on a manifold $N$ with $\dim N = \dim M+1$. In the present paper, we also require that $f: N \to X$ is an embedding. 

\item
It can happen that $f : N \to X$ is topologically nontrivial in the sense of bordism, but the homology class $f_* [N]$ is zero. To have a consistent shifted quantization condition, we need the anomaly evaluated on such an $N$ to be trivial (or more precisely it is completely determined by the anomaly polynomial of perturbative anomaly).

\end{enumerate}
We will discuss these points in Section~\ref{sec:Homsub} for the case relevant to the RR 5-form $G_5$ in Type~IIB string theory. Let us neglect these issues in this section.

\subsection{General structure}

Neglecting the above issues, we can discuss quite generally how to determine the shifted quantization condition. Let $M$ be a worldvolume of a brane. We denote the dimension of $M$ as $p-1=\dim M$. When the worldvolume theory has an anomaly, we take a manifold $N'$ whose boundary is $\partial N' = M$. The dimension of $N'$ is $\dim N' =p$. By considering some appropriate term on $N'$, it is expected\,\footnote{Nonperturbative anomaly inflow is described for fermions in \cite{Witten:1999eg,Witten:2015aba,Witten:2019bou} and for $p$-form fields in \cite{Hsieh:2020jpj}. These cases are sufficient for our purposes in the present paper. } that the anomaly inflow mechanism makes the partition function of the worldvolume theory well-defined. We denote the partition function by $\cZ_\textrm{matter}(N')$. If we use another manifold $N''$, we get another value for the partition function $\cZ_\textrm{matter}(N'')$. Their ratio is given by
\beq
\frac{\cZ_\textrm{matter}(N'')}{\cZ_\textrm{matter}(N')}  = \cZ_\textrm{anomaly}(N) \in \U(1), \qquad N = N'' \cup \overline{N'}
\eeq
where $ \cZ_\textrm{anomaly}(N)$ is a $\U(1)$-valued partition function representing the anomaly. 

It is not at all an essential restriction on $N$ that the manifold $N$ is constructed by gluing $N'$ and $N''$. For instance, if we are given an arbitrary closed manifold $N$, we take a small disk $D^{p} \subset N$. Then $N$ can be regarded as obtained from gluing $D^{p}$ and $N \setminus D^{p}$ along the common boundary $\partial D^{p} = S^{p -1}$. More easily (but somewhat abstractly), we may just consider the case $M = \varnothing, N'=N$ and $N'' = \varnothing$. Therefore, $N$ can be any manifold as long as it has some appropriate structure (such as some kind of orientation) for the worldvolume theory of a brane, including the anomaly inflow term, to make sense. The relevant structure type for the case of a D3-brane will be discussed in Section~\ref{sec:D3structure}.

Anomalies have a further property. There is a polynomial $\sA_{p+1}$ of fields in $X$, which we call an anomaly polynomial, with the following properties. The $\sA_{p+1}$ is a closed differential $(p+1)$-form on $X$. Suppose that $N$ is a boundary of some manifold $L$, $N =\partial  L$. Then $\cZ_\textrm{anomaly}(N)$ is given by
\beq
\cZ_\textrm{anomaly}(\partial L) = \exp\left( 2\pi \i \int_{L} \sA_{p+1} \right), \label{eq:compatibility}
\eeq
where we used the fact that $\dim L=p+1$ so that we can integrate the differential $(p+1)$-form $ \sA_{p+1}$.
We will give the explicit $\sA_{p+1}$ in the case of the anomaly of a D3-brane in Section~\ref{sec:perturbative}. Notice that $\cZ_\textrm{anomaly}(N) $ depends only on $N$, so the right hand side of \eqref{eq:compatibility} should be independent of a choice of $L$. This implies that if $L$ is a closed manifold, we have
\beq
 \int_{L} \sA_{p+1}  \in \bZ \quad \text{if $\partial L = \varnothing$}.
\eeq

Let us summarize the situation so far. The anomaly of a brane is represented by a map which assigns a $\U(1)$ value for each closed manifold $N$,
\beq
 \cZ_\textrm{anomaly} : N \mapsto  \cZ_\textrm{anomaly}(N) \in \U(1).
\eeq
Moreover, it has the property \eqref{eq:compatibility}. These are the properties that characterize a ``$p$-form field'', although  it is not an independent field but is determined by other fields on $X$ as a composite. Roughly, we may write 
\beq
\cZ_\textrm{anomaly}(N) = \exp\left(2\pi \i \int_N \sB_p \right), \qquad \sA_{p+1}= \d \sB_p. \label{eq:p-schematic}
\eeq
However, in general $\sB_p$ is not necessarily defined as a differential form. The obstruction for $\sB_p$ to being a differential form is measured by the cohomology class of the ``$p$-form field''. In general, the relevant cohomology theory describing $\cZ_\textrm{anomaly}$ may be a generalized cohomology theory~\cite{Yamashita:2021cao} known as the Anderson dual of bordism homology. (See \cite{Freed:2016rqq} for the original conjecture about the relevance of the Anderson dual of bordism homology). However, for our applications in the present paper, it turns out that we only need ordinary cohomology $H^{p+1}(X; \wt \bZ)$ possibly with a twisted coefficient. Then the topology of ``$p$-form fields'' is classified by $H^{p+1}(X; \wt \bZ)$. We obtain an element 
\beq
\sC_{p+1} \in H^{p+1}(X; \wt \bZ) 
\eeq
that is the topological class of $\cZ_\textrm{anomaly}$. It has the property that its image in the real cohomology group $H^{p+1}(X; \wt \bR) $, which we denote as $[\sC_{p+1}]_\bR$, is the de~Rham cohomology class $[\sA_{p+1}]$ of  $\sA_{p+1}$,
\beq
[\sC_{p+1}]_\bR=[\sA_{p+1}] \in H^{p+1}(X; \wt \bR) .
\eeq 
In general, $\sC_{p+1}$ contains more information than just the de~Rham class $[\sA_{p+1}]$. For example, if $\sA_{p+1}=0$, then $\sB_p$ should be regarded as an element of $H^{p}(X; \wt {\bR/\bZ})$, and $\sC_{p+1}$ is given in terms of the Bockstein homomorphism $\wt \beta : H^{p}(X; \wt {\bR/\bZ}) \to H^{p+1}(X; \wt {\bZ}) $ as 
\beq
\sC_{p+1} = \wt \beta ( \sB_p).
\eeq
On the other hand, if $\sC_{p+1}=0$ (but with possibly nonzero $\sA_{p+1}$), the ``$p$-form field'' is topologically trivial, and $\sB_{p}$ is really a differential $p$-form such that $\d \sB_p = \sA_{p+1}$.

More precise definition of ``$p$-form fields'' is given in terms of (generalized) differential cohomology theory. In particular, $(\sA_{p+1}, \sB_p, \sC_{p+1})$ gives an element of the differential cohomology group $\widehat H^{p+1}(X; \wt \bZ)$ (see \cite{CS,Hopkins:2002rd,Freed:2006yc,Hsieh:2020jpj,Yamashita:2021cao}).

Now we can discuss the shifted quantization condition and the obstruction to its existence. Suppose that the brane under consideration is coupled to a $(p-1)$-form field whose field strength $p$-form is denoted as $G_p$. Then the total partition function of the brane is given by
\beq
\cZ(N') : = \cZ_\textrm{matter}(N') \exp\left( 2\pi \i \int_{N'} G_p \right).
\eeq
We want it to be independent of a choice of $N'$. This requires
\beq
 \exp\left( -2\pi \i \int_{N} G_p \right)= \cZ_\textrm{anomaly} (N) \label{eq:shiftedcondition}
\eeq
for any closed manifold $N$ (with appropriate structure). 

From the point of view of $G_p$, the condition \eqref{eq:shiftedcondition} gives a shifted quantization condition because it implies
\beq
 \int_{N} G_p \in \frac{\i}{2\pi} \log  \cZ_\textrm{anomaly} (N)  + \bZ.
\eeq
On the other hand, from the point of view of $\cZ_\textrm{anomaly}$, it implies that the $\sB_p$ in \eqref{eq:p-schematic} is given by 
\beq
\sB_p = - G_p.
\eeq
Therefore, $\sB_p$ is a differential $p$-form. This fact implies that the cohomology class $\sC_{p+1}$ must be zero for the shifted quantization condition to be consistent. In other words, $\sC_{p+1}$ gives the obstruction to the existence of a shifted quantization. Alternatively, as discussed in Section~\ref{sec:Mtheory} and \ref{sec:example1}, we may add codimension-$(p+1)$ branes on the Poincare dual of $-\sC_{p+1}$ to satisfy the Gauss law constraint for $G_p$.

If $\sC_{p+1}$ is trivial, we have
\beq
\d G_p =  - \sA_{p+1}.\label{eq:Bianchi}
\eeq
This is the modified Bianchi identity for $G_p$. This has a clear meaning in supergravity as we discuss in Section~\ref{sec:perturbative}.

A $(p-1)$-form field $C_{p-1}$ whose field strength is $G_p$ has more information than just the field strength $G_p$. In a mathematical terminology, we may say that the $(p-1)$-form field is a ``trivialization'' of the anomaly $p$-form field. We omit the details.

\section{Some details on topology and geometry}\label{sec:Homsub}

In this section, we discuss the structure type needed for the worldvolume of a D3-brane, and the subtleties mentioned at the beginning of Section~\ref{sec:anomalycohomology}. Some of these subtleties are variants of the questions studied in the seminal paper by Thom~\cite{Thom:MR0061823} on (co)bordism.

This section is somewhat technical and may be skipped if the reader accepts that the discussions in Section~\ref{sec:anomalycohomology} work for the D3-brane. Some of the results of this section may be summarized as follows. Let $f : N \to X$ be a map from a manifold $N$ such that we want to evaluate the anomaly $\cZ_\textrm{anomaly} (N) $ on $N$ as in Section~\ref{sec:anomalycohomology}.
\begin{itemize}

\item The structure needed for $f: N \to X$ is that its homology class $[N]$ (or more precisely $f_* [N]$) takes values in $H^5(X; \wt \bZ)$, where $\wt \bZ$ is twisted by the bundle $\zeta$ associated to the $\GL(2,\bZ)$ bundle by taking determinant. We do not require other conditions on $N$. (In particular, we do not require $N$ to admit a $\spin^c$-structure.)

\item Any homology class in $H^5(X; \wt \bZ)$ can be represented by a map $f : N \to X$ as $f_*[N]$. Moreover, $N$ can be taken to be a submanifold in the spacetime $X$, i.e. $f : N \to X$ can be taken to be an embedding.

\item Roughly speaking, a manifold $L$ such that $\partial L = N$ (when $f : N \to X$ admits such $L$) can be taken to be an immersion such that the equation \eqref{eq:compatibility} works. More detailed results are stated and used in Section~\ref{sec:perturbative}.

\item Nontrivial bordisms $f : N \to X$ such that $f_*[N]$ is zero will be treated in Section~\ref{sec:trivialhomology}.
\end{itemize}

\subsection{The structure type of D3-brane worldvolume manifolds}\label{sec:D3structure}
Consider a worldvolume $M$ of a D3-brane. 
If we change the orientation of $M$, a D3-brane becomes an anti-D3-brane. Therefore, one might think that $M$ must be oriented. However, in the presence of a general $\GL(2,\bZ)$ duality bundle, the condition for orientation is modified as follows. Elements such as $\Omega, (-1)^{\sF_L} \in \GL(2,\bZ)$ change the sign of the Ramond-Ramond 4-form $C_4$ and hence they change a D3-brane to an anti-D3-brane. Let $\zeta$ be the $\bZ_2$ bundle on $X$ associated to the $\GL(2,\bZ)$ duality bundle that is obtained by taking the determinant of $\GL(2,\bZ)$. If the orientation bundle of $M$ is the same as the $\zeta$ restricted to $M$, then the worldvolume makes sense as a D3-brane. In particular, the homology class $[M]$ of $M$ in $X$ takes values in $H_4(X, \wt \bZ)$ where $\wt \bZ$ is twisted by $\zeta$ as in \eqref{eq:twistZ}.

In fact, it is convenient to regard the D3-brane worldvolume theory to be the dimensional reduction of the M5-brane worldvolume theory compactified on $T^2$.  The $\GL(2,\bZ)$ acts on $T^2$. Then the above condition on orientation just means that the worldvolume of the M5-brane is orientable (and we need to orient it to distinguish a D3-brane from an anti-D3-brane).

It is not necessary to require that the worldvolume $M$ is a spin manifold for the following reason. Let $\cV$ be the rank 2 real vector bundle associated to the $\GL(2,\bZ)$ bundle. 
The conditions needed for the spacetime $X$ of Type~IIB string theory (or F-theory without singular fibers) are that the rank 12 bundle $TX \oplus \cV$ has a spin, or more precisely $\pin^{+}$, structure, and $X$ itself is oriented. The former condition follows from the duality between F-theory and M-theory, and the latter condition follows from the fact that the fields of Type~IIB string theory are chiral. The fermions on a D3-brane are sections of the $\pin^{+}$ bundle associated to the restriction of $TX \oplus \cV$ to $M$,
\beq
(TX \oplus \cV)|_M = TM \oplus \nu(M) \oplus \cV|_M, \label{eq:TMNV}
\eeq
where $\nu(M)$ is the normal bundle to $M$. We do not need spin structures for each of $TM$, $ \nu(M)$, and $ \cV|_M$ separately. The $\pin^{+}$ structure of the total bundle \eqref{eq:TMNV} is enough for the D3-brane worldvolume theory to make sense.\footnote{At least this is the case for the fermions. The $\U(1)$ gauge field on a D3-brane with this most general situation is not yet studied. See \cite{Hsieh:2020jpj} for the study of a case in which $\nu(M)$ is trivial and the duality bundle is $\SL(2,\bZ) \subset \GL(2,\bZ)$. The author believes that the claim for the most general case is correct. } 

We also do not require that the normal bundle $\nu(M)$ admits a $\spin^c$ structure (or its $\pin$ generalization), which is the usual requirement of the K-theoretic interpretation of D-branes~\cite{Witten:1998cd}. The reason is as follows. First of all, if we include a $B$-field background, then the condition is modified~\cite{Witten:1998xy,Witten:1998cd,Freed:1999vc}. More fundamentally, however, the worldvolume theory of a D-brane makes sense, at least as a quantum field theory, even if $\nu(M)$ does not admit a $\spin^c$ structure and there is no $B$ field background. For simplicity of discussion, let us consider the case that the $\GL(2,\bZ)$ bundle is trivial. (See \cite{Hsieh:2019iba,Hsieh:2020jpj} for the case of $\SL(2,\bZ)$ backgrounds.) The $\U(1)$ gauge field on a D-brane is a $\spin^c$ connection of the normal bundle $\nu(M)$. For the trivial $\GL(2,\bZ)$ bundle, the second Stiefel-Whitney class of the tangent bundle $w_2(M)$ is the same as that of the normal bundle. In terms of the field strength 2-form $F$ of the $\U(1)$ gauge field, the condition that the $\U(1)$ gauge field is a $\spin^c$ connection is described by a shifted quantization condition which is roughly written as
\beq
[F] \in \frac{1}{2} w_2( M) + H^2(M; \bZ). \label{eq:spinc}
\eeq
Let $\beta : H^2(M; \bR/\bZ) \to H^3(M; \bZ)$ be the Bockstein homomorphism associated to the short exact sequence $0 \to \bZ \to \bR \to \bR/\bZ \to 0$. We regard $\frac{1}{2} w_2( M)$ as an element of $H^2(M; \bR/\bZ)$ and denote it as $w'_2(M) \in H^2(M; \bR/\bZ)$. Then we define 
\beq
W_3( M) = \beta\left( w'_2(M) \right).
\eeq
As in the discussions of Section~\ref{sec:Minc}, the condition \eqref{eq:spinc} gives an equation for $W_3$ which is roughly written as
\beq
\d F = W_3( M).
\eeq
(More precise discussions may be done in terms of differential cochains.)
The left hand side is an ``exact form'', so the right hand side should be topologically trivial. The vanishing of the cohomology class $W_3( M)$ is precisely the condition for the existence of a $\spin^c$ connection. However, suppose we include a 't~Hooft operator along a codimension-3 submanifold $C \subset M$, and let $\delta_C$ be the delta function 3-form supported on $C$. Then the equation is modified to
\beq
\d F = W_3(M) + \delta_C.
\eeq
Thus, by taking $C$ to be (the negative of) the Poincare dual of $W_3(M) $, we can solve the equation for $F$. 

As a quantum field theory, the above situation is conceptually not so different from the case of e.g., massless free fermions in 4-dimensions coupled to a background gauge field of some gauge group. If we consider an instanton configuration of the background gauge field, the partition function is zero due to fermion zero modes. This does not mean that such an instanton background is inconsistent. If we insert some fermion operators, correlation functions can be nonzero. In conceptually the same way, the partition function of the $\U(1)$ gauge field on a D-brane is zero if the background has nonzero $W_3( M) $, but correlation functions including 't~Hooft operators can be nonzero. 

In the actual string theory as opposed to just a quantum field theory, a 't~Hooft operator on a D$p$-brane is realized by letting a D$(p-2)$-brane end on the D$p$-brane along a codimension-3 submanifold $C$. This is consistent with K-theory, since the D$(p-2)$-brane modifies the K-theory class of the brane configuration. As far as anomalies on the D$p$-brane is concerned, ('t~Hooft) operator insertion is irrelevant and we can consider anomalies even if $ W_3( M)  \neq 0$. 

In summary, the only structure we require for the worldvolume of a D3-brane $M$ is that its orientation bundle is correlated with the $\bZ_2$ bundle $\zeta$ associated to the $\GL(2,\bZ)$ duality bundle. For the study of anomalies, we consider manifolds $N$ of dimension $\dim N = \dim M+1=5$ with the same structure as $M$, because the term on $N$ used in the anomaly inflow requires the same structure as in $M$. In particular, the homology class $[N]$ takes values in $H_5(X; \wt \bZ)$,
\beq
[N] \in H_5(X; \wt \bZ),
\eeq
where the coefficient system $\wt \bZ$ is twisted by $\zeta$. 

\subsection{Homology and bordism}\label{sec:hombom}

In this subsection we study whether any element of $ H^5(X; \wt \bZ)$ can be realized by a map $ f : N \to X$ as $f_*[N]$, where $[N]$ is the fundamental homology class of $N$.\footnote{We are going to use some knowledge of Thom isomorphism, Pontryagin-Thom construction, bordism, and Atiyah-Hirzebruch spectral sequence. See e.g. \cite{MilnorStasheff:MR0440554,Hatcher2,Garcia-Etxebarria:2018ajm,FreedLectures} for some exposition. }
The twisted coefficient may be treated as follows. Let $\wt \bR$ be the real rank 1 bundle on $X$ twisted by $\zeta$, and let ${\rm Thom}(\wt \bR)$ be the Thom space. By Thom isomorphism we have 
\beq
 H_i(X; \wt \bZ)= \wt H_{i+1}({\rm Thom}(\wt \bR);  \bZ)  ,
\eeq
where $\wt H$ is the reduced homology and the isomorphism is given by the cap product with the Thom class.

Let $\Omega^{\rm SO}_{i, \zeta}$ be the bordism group that consists of equivalence classes $[f:N \to X]$ of maps $ f : N \to X$ for $\dim N = i$ such that the orientation bundle of $N$ is isomorphic to the pullback $f^* \zeta$ of $\zeta$. By Pontryagin-Thom construction, this bordism group is given by
\beq
\Omega^{\rm SO}_{i, \zeta}(X) = \wt \Omega^{\rm SO}_{i+1}({\rm Thom}(\wt \bR)),
\eeq
where $\wt \Omega^{\rm SO}$ is the reduced oriented bordism. 

There is a natural map
\beq
\Omega^{\rm SO}_{i, \zeta}(X)  \to H_5(X; \wt \bZ) \label{eq:bordismtohomology}
\eeq
given by $[f:N \to X] \mapsto f_* [N]$.
We want to show that this map is surjective.

The oriented bordism groups of a point ${\rm pt}$ are known to be
\beq
\Omega^{\rm SO}_0({\rm pt}) \simeq \bZ, \quad  \Omega^{\rm SO}_4({\rm pt}) \simeq \bZ, \quad \Omega^{\rm SO}_5({\rm pt}) \simeq \bZ_2, \quad \Omega^{\rm SO}_i({\rm pt})=0~~(i=1,2,3,6,7). \label{eq:pbordism}
\eeq
The second page of the Atiyah-Hirzebruch spectral sequence for $\wt \Omega^{\rm SO}_{i+1}({\rm Thom}(\wt \bR))$ is given by 
\beq
E^2_{i+1,j}=\wt H_{i+1} ({\rm Thom}(\wt \bR); \Omega^{\rm SO}_j({\rm pt}) ) = H_{i} (X; \Omega^{\rm SO}_j({\rm pt}) \otimes_{\bZ} \wt \bZ) .
\eeq
We are interested in whether the term 
\beq
E^2_{6,0} =H_{5} (X;  \wt \bZ)
\eeq
survives or not when we go to the infinite page $E^\infty_{6,0}$.  If it survives, it implies that any element of $H_{5} (X;  \wt \bZ)$ can be realized by some $[f : N \to X]$. The only possibility for nonzero differential involving $E^\bullet_{6,0}$ is $d_5 : E^5_{6,0} \to E^5_{1,4}$ which is
\beq
d_5 : H_{5} (X;  \wt \bZ) \to H_{0} (X;  \Omega^{\rm SO}_4({\rm pt}) \otimes_{\bZ} \wt \bZ).
\eeq
However, this differential must be zero because elements of $E_{1,4}=H_{0} (X;  \Omega^{\rm SO}_4({\rm pt}) \otimes_{\bZ} \wt \bZ)$ must survive by the following reason. The generator of $\Omega^{\rm SO}_4({\rm pt})$ is represented by the complex projective space $\mathbb{CP}^2$. The generator of $H_{0} (X;  \Omega^{\rm SO}_4({\rm pt}) \otimes_{\bZ} \wt \bZ) $ is represented by a constant map $\mathbb{CP}^2 \to \{ {\rm point} \} \subset X$.
If the bundle $\wt \bR$ is trivial and $X$ is connected, we have $H_{0} (X;  \Omega^{\rm SO}_4({\rm pt}) \otimes_{\bZ} \wt \bZ) \simeq \bZ$. In this case, $\mathbb{CP}^2$ is oriented and we can detect $\mathbb{CP}^2$ by the Pontryagin number $\int_{\mathbb{CP}^2} p_1$ where $p_1$ is the first Pontryagin class. If the bundle $\wt \bR$ is nontrivial and $X$ is conntected, we have $H_{0} (X;  \Omega^{\rm SO}_4({\rm pt}) \otimes_{\bZ} \wt \bZ) \simeq \bZ_2$. In this case, we can detect $\mathbb{CP}^2$ by the Stiefel-Whitney number $\int_{\mathbb{CP}^2} w_4$ which is nonzero because of the relation \eqref{eq:CWS} and the fact that the total Chern class of the complex bundle $T\mathbb{CP}^2$ is given by $(1+\sc)^3$, where $\sc \in H^2(\mathbb{CP}^2;\bZ)$ is a generator. Thus, in both cases, $\mathbb{CP}^2 \to \{ {\rm point} \} \subset X$ is nontrivial. We conclude that $H_{0} (X;  \Omega^{\rm SO}_4({\rm pt}) \otimes_{\bZ} \wt \bZ)$ and hence $H_{5} (X;  \wt \bZ)$ survive to the infinite page.

\subsection{Elements of bordism groups that are trivial in ordinary homology}\label{sec:bortohom}

We have seen in Section~\ref{sec:hombom} that any element of $H_5(X; \wt \bZ)$ is realized by an element of $\Omega^{\rm SO}_{5, \zeta}(X) $. In other words, the natural map \eqref{eq:bordismtohomology} is surjective. 

There are elements of $\Omega^{\rm SO}_{5, \zeta}(X) $ that map to zero in $H_5(X; \wt \bZ)$. In the Atiyah-Hirzebruch spectral sequence, the second page contains nonzero terms
\beq
E^2_{1,5}  = H_{0} (X;  \Omega^{\rm SO}_5({\rm pt}) \otimes_{\bZ} \wt \bZ), \qquad  E^2_{2,4}  = H_{1} (X;  \Omega^{\rm SO}_4({\rm pt}) \otimes_{\bZ} \wt \bZ).
\eeq
Some of them may survive to the infinite page. The elements of bordisms corresponding to them may be constructed as follows. For $H_{0} (X;  \Omega^{\rm SO}_5({\rm pt}) \otimes_{\bZ} \wt \bZ)$, we take a neighborhood $D^{10}$ of an arbitrary point in $X$, where $D^n$ is an $n$-dimensional disk. Then we consider a map $f : Z \to D^{10} \subset X$ where $Z$ is the generator of $ \Omega^{\rm SO}_5({\rm pt})$. (The generator will be discussed in Section~\ref{sec:ex2}.) For $H_{1} (X;  \Omega^{\rm SO}_4({\rm pt}) \otimes_{\bZ} \wt \bZ)$, we first choose a circle $S^1 \subset X$ representing an element of $H_{1} (X;  \wt \bZ)$, and take its tubular neighborhood $S^1 \times D^9 \subset X$. Then we consider a map $ f : S^1 \times \mathbb{CP}^2 \to S^1 \times D^9 \subset X$ that consists of the product of two maps $S^1 \to S^1$ and $\mathbb{CP}^2 \to D^9$, where the part $S^1 \to S^1$ is the identity, and $\mathbb{CP}^2$ is the generator of $\Omega^{\rm SO}_4({\rm pt})$.

In principle, the anomaly of a D3-brane can be nonzero on such manifolds. In that case, there is no hope to cancel the anomaly by using fluxes of $G_5$. In Section~\ref{sec:trivialhomology}, we will check that the anomaly is zero on them. It is realized by nontrivial cancellations between the anomalies of the fermions and the $\U(1)$ gauge field on a D3-brane.

\subsection{Embedding of manifolds}
We have seen that any element of $H_{5} (X;  \wt \bZ)$ is represented by a map $f : N \to X$. Now we show that the map can be taken to be an embedding without changing the bordism class. First notice that
\beq
\dim X = 10 = 2\dim N.
\eeq
In this case, Whitney immersion theorem says that any map $f : N \to X$ is homotopic to an immersion such that the inverse image $f^{-1}(p)$ of a point $p \in X$ contains at most two points, and if it contains two points $p_1$ and $p_2$, then the neighborhoods of $p_1$ and $p_2$ intersect transversely. Such an immersion is achieved just by taking the map $f : N \to X$ generic enough (as we briefly review around \eqref{eq:singular}). 

Suppose that $p \in X$ is such a self-intersection point. Take a local coordinate system $x^i~(i=1,\cdots,10)$ around $p \in X$ such that $p$ corresponds to $x^i=0$. We denote
\beq
\vec y =(x^1,x^2,x^3,x^4,x^5)^{\rm T}, \qquad \vec y\,' = (x^6, x^7, x^8, x^9, x^{10})^{\rm T}.
\eeq
 In an appropriate coordinate system, $N$ is immersed in $X$ as
\beq
\vec y=0~ \text{ or } ~\vec y\,'=0. \label{eq:localN}
\eeq
Then we replace $N$ by another manifold $N'$ such that it is embedded in $X$ as
\beq
\frac{\vec y}{|\vec y|} = \frac{\vec y\,'}{|\vec y\,'|} , \qquad |\vec y| |\vec y\,'| = \epsilon, \label{eq:localN'}
\eeq
where $\epsilon>0$ is very small. This is a local modification near the point $p$ and does not change the region far from it. As a manifold, $N'$ is obtained from $N$ by eliminating neighborhoods $D^{10}$ of the points $p_1, p_2 \in f^{-1}(p)$ and connecting the boundaries $S^9$ by a tube of the form $[0,1] \times S^9$.\footnote{See e.g., \cite{Milnor} for an exposition of the surgery used here.}  Then we can take a map $f' : N' \to X$ such that the embedding \eqref{eq:localN'} is realized.

The new configuration is bordant to the original configuration. To see it, we take a manifold $L$ whose boundary is $N' \sqcup \overline{N}$ where $\sqcup$ is the disjoint union. The $L$ is locally described by coordinates $(\vec z, w, t)  \in \bR^5 \times \bR \times [-\epsilon,\epsilon]$ with
\beq
 \vec z\,^2 - w^2 =2t, \quad t \in [-\epsilon, \epsilon]. \label{eq:Morse}
\eeq
The boundaries $t= \epsilon$ and $t =-\epsilon$ correspond to $N'$ and $N$, respectively. We take a map  $g: L \to X$ given by
\beq
\vec y = \left( \frac{\sqrt{\vec z\,^2 + 2 \epsilon} + w }{2\sqrt{\vec z\,^2 +  (\epsilon-t)}} \right)  \vec z, \quad \vec y\,' =   \left( \frac{\sqrt{\vec z\,^2 + 2\epsilon} - w}{2\sqrt{\vec z\,^2 +  (\epsilon-t)}} \right)  \vec z. \label{eq:resolutionbordism}
\eeq
One can check that it gives an explicit bordism from \eqref{eq:localN} at $t=-\epsilon$ to \eqref{eq:localN'} at $t = \epsilon$. Therefore, the bordism classes of $f : N \to X$ and $f' : N' \to X$ are the same, and the new map $f': N' \to X$ is an embedding in this local region. We can perform this operation for each self-intersection point to obtain an embedding.

\subsection{Bordism between embedded manifolds}\label{sec:betweenN0N1}
Our next task is to study the following question. Suppose we have two submanifolds $N_0$ and $N_1$ (embedded in $X$) that are bordant by a map (not necessarily embedding) $g : L \to X$. The boundaries of $L$ are $\partial L = N_1 \sqcup \overline{N_0}$. The situation would have been simple if we could have taken $g : L \to X$ to be an embedding. However, this is not possible in general. For instance, if $N_0$ and $N_1$ intersect with each other in $X$, there is no hope for $L$ to be an embedding. What nice properties can we assume about this bordism?

The essential property we want to require about $L$ is that the normal bundle to $L$ is well-defined and has at most the same rank as the normal bundle $\nu(M)$ to $M$ in $X$, where $M$ is a world volume of a D3-brane and hence the rank of $\nu(M)$ is 6. From the worldvolume point of view, the normal bundle is regarded as a ``gauge bundle'' whose Lie algebra is $\so({\rm rank}(\nu))$, where ${\rm rank}(\nu)$ is the rank of the normal bundle. We need ${\rm rank}(\nu)$ to be the same or smaller when we go from $M$ to higher dimensional manifolds as in Section~\ref{sec:anomalycohomology} for the discussions of anomaly inflow in that section to make sense, because the ``gauge group'' should not become larger.\footnote{See \cite{Witten:2019bou} for general discussions of how to construct the anomaly $\cZ_{\rm anomaly}(N)$ and the anomaly polynomial $\sA_{p+1}$ by starting from an anomalous theory for the case of fermions.} There is no problem for the rank to be smaller, as in the case of an embedding of $N$ in $X$ (with $\dim N = \dim M +1$) in which case we have ${\rm rank}(\nu(N)) = {\rm rank}(\nu(M)) -1$; a theory with the symmetry $\so({\rm rank}(\nu))$ can be coupled to background fields for the smaller symmetry $\so({\rm rank}(\nu)-1)$.

For the purpose of the previous paragraph, we can relax the conditions on a bordism $L$ as follows. First, instead of the 10-manifold $X$, we consider 
\beq
\bR \times [0,1]  \times X .
\eeq
Then we take $N_0$ and $N_1$ to be embedded in
\beq
&N_i \subset \{0 \} \times \{ i \}   \times X \subset \bR \times [0,1]  \times X, \qquad (i=0,1).
\eeq 
The bordism is taken to be a map
\beq
g: L \to \bR \times [0,1]  \times X.
\eeq 
We will require $L$ to be an immersion rather than an embedding, since that is enough for the normal bundle $\nu(L)$ to be well-defined.\footnote{It is important to require that $N_0$ and $N_1$ are embeddings rather than immersions as we explain later. }  We will also require that $L$ intersects transversally with $\bR \times \{i\} \times X$~($i=0,1$) along $N_i$. These properties are enough for the normal bundle to have the desired properties. It has the same rank $6$ as the normal bundle of a D3-brane worldvolume in $X$. The restriction of the normal bundle to $\partial L = N_1 \sqcup \overline{N_0}$ is the normal bundle of $N_i$ in $X$ with a trivial bundle $\underline{\bR}$ added,
\beq
\nu(L)|_{N_i} = \nu(N_i) \oplus \underline{\bR},
\eeq
where $\nu(L)$ is the normal bundle in $\bR \times [0,1]  \times X$, and $\nu(N_i)$ is the normal bundle in $X$. We remark that if $N_0$ and $N_1$ are bordant in $X$, then they are bordant also in $\bR \times [0,1] \times X$.

The dimensions are such that $\dim (\bR \times [0,1]   \times X) = 12 = 2\dim L$. Thus a map $g: L \to \bR \times [0,1]  \times X$ becomes an immersion simply by taking $g$ generic enough. However, for our later computation it will turn out that we need more information, so we study it in more detail. 

Let us ask whether we can take a map $g$ such that its image is contained in $ [0,1] \times X$ as
\beq
g : L \to [0,1]  \times X \simeq \{0\}  \times [0,1]  \times X \subset \bR  \times [0,1]  \times X.
\eeq
Here we identify $ [0,1]  \times X$ with $\{0\}  \times [0,1]  \times X$. Of course such a map exists, but it is not guaranteed to be an immersion even if we take $g$ to be generic enough. 

To study this point, consider a more general case 
\beq
g: Z \to Y, \qquad \dim Z = n, \qquad \dim Y = m, \qquad (n < m) \label{eq:genericmap}
\eeq
where $Z$ and $Y$ are manifolds, and $Z$ is assumed to be compact (possibly with boundaries). 
A map $g$ is an immersion if its Jacobian matrix 
\beq
\d g _p: T_p Z \to T_{g(p)} Y
\eeq 
has the maximum rank $n$ for all points $p \in Z$. 

Let $\bM (m,n\,;r)$ be the space of $m \times n$ matrices with rank $r$. Its dimension is $\dim \bM(m,n\,;r) = r(n+m-r)$ because, after permuting some rows and columns if necessary, elements of $\bM (m,n\,;r)$ are of the form 
\beq
\begin{pmatrix} A & B \\ C & C A^{-1} B \end{pmatrix} = \begin{pmatrix} A  \\ C &\end{pmatrix} \begin{pmatrix} I_r & A^{-1}B \ \end{pmatrix}
\eeq
where $A$ is an $r \times r$ matrix with $\det A \neq 0$, and $I_r$ is the $r \times r$ unit matrix. Thus the codimension of $\bM (m,n\,;n-1)$ (i.e. rank $n-1$ matrices) inside $\bM (m,n\,;n)$ (i.e. the maximum rank matrices) is $nm-(n-1)(m+1) = m-n+1$. If the map \eqref{eq:genericmap} is generic enough, we expect that the Jacobian matrix $\d g_p$ becomes rank $n-1$ on a subspace $S \subset Z$ of dimension 
\beq
\dim S = n-(m-n+1) = 2n-m-1. \label{eq:singular}
\eeq
When $\dim S<0$, we expect that the Jacobian matrix always has the maximum rank for generic enough $g$. This is the essential reason for the Whitney immersion theorem in the case $m=2n$. On the other hand, when $m=2n-1$ as in the case of $(n,m) = (6,11)$, we have $\dim S=0$ and hence $S$ consists of 0-dimensional points for generic enough $g$.\footnote{More precisely, Sard theorem may be used to show that if we make arbitrarily small deformations of any $g$, then points of $S$ are isolated. } For generic enough $g$, there are no points at which the rank of $\d g_p$ is less than $n-1$.

Locally,\footnote{Local properties are the same as in \cite{Whitneyimmersion} which studies the case $Y = \bR^{2n-1}$.} 
by choosing appropriate coordinate systems $\vec z  = (z_1, \cdots, z_n)^{\rm T} $ for $Z$ and $\vec y = (y_1, \cdots, y_m)^{\rm T}$ for $Y$, the Jacobian matrix near a point $p \in S$ (where $p$ is at $\vec z=0$ and $g(p)$ is at $\vec y =0$) is given by
\beq
\d g  =  \begin{pmatrix} I_{n-1} & 0 \\ 0 & \vec z \end{pmatrix}. \label{eq:Jsing}
\eeq
This Jacobian matrix describes the behavior near a generic isolated point at which the rank is reduced to $n-1$.
Let us denote $\vec z\,' = (z_1, \cdots, z_{n-1})^{\rm T}$. The map $g$ itself near $p$ is
\beq
g = \begin{pmatrix} \vec z\,' \\  z_n \vec z\,'   \\ \frac{1}{2} z_n^2  \end{pmatrix}. 
\eeq
Notice that two points $\vec z = (0,\cdots,0, \pm z_n)^{\rm T}$ with $z_n \neq 0$ are mapped to the same point on $Y$, 
\beq
g (0,\cdots,0, \pm z_n) =  \begin{pmatrix}  0 \\ \vdots \\ 0 \\ \frac12 z_n^2   \end{pmatrix}. 
\eeq
Thus the point $g(p) \in Y$ (which is $\vec y =0$) is an endpoint of a curve $\ell \subset Y$ along which $Z$ intersects with itself. In dimensions $n \geq 3$, we may not worry about intersections of lines for generic enough $g$. There are no points which are more singular than \eqref{eq:Jsing}. 

If $Z$ is a closed manifold, the intersection curve $\ell$ may have a finite length. Then, $\ell$ must have two endpoints $g(p), g(q)$ for some other point $q \in Z$. In other words, points of $S$ appear in pairs $(p,q)$ such that $g(p)$ and $g(q)$ are connected by a curve $\ell \subset Y$. 

Let us return to the case $Z = L$ and $Y =  [0,1] \times X$. If $\ell$ could intersect with the boundaries $  \{i\} \times X$ (where $i=0,1$), we would have the situation that one of the endpoints $g(q)$ is ``outside the boundaries''. However, recall that $N_0$ and $N_1$ are assumed to be embeddings and hence $\ell$ cannot intersect with the boundaries. Therefore, a curve $\ell$ must be contained in the interior $ (0,1)\times X$. In particular, the number of points in $S$ is even. 

We emphasize that the above result is not guaranteed to hold if $N_0$ or $N_1$ is not an embedding. In fact, the bordism \eqref{eq:resolutionbordism} essentially gives an explicit example in which $S$ contains only a single point and $\ell$ intersects with one of the boundaries. More precisely, we add one coordinate $ u \in  [0,1]$ to $X$, and define the map $g : L \to [0,1] \times X$ as
\beq
\vec y = \left( \frac{\sqrt{\vec z\,^2 + 2 \epsilon} + w }{2\sqrt{\vec z\,^2 +  (\epsilon-t)}} \right)  \vec z, \quad \vec y\,' =   \left( \frac{\sqrt{\vec z\,^2 + 2\epsilon} - w}{2\sqrt{\vec z\,^2 +  (\epsilon-t)}} \right)  \vec z, \quad u =\frac{t+\epsilon}{2\epsilon}, \label{eq:singlesingular}
\eeq
where $\vec z, w, t$ are related by \eqref{eq:Morse}.
One can see that the singular point $p$ is at $(\vec z, w)=(0,0)$, its image $g(p)$ is at $(\vec y, \vec y\,', u) = (0,0, \frac12 )$, and the curve $\ell$ is given by $\vec y = \vec y\,' =0$ and $u \leq \frac12$ corresponding to $\vec z=0$ and $t = - \frac12 w^2$. 

Now let us add $\bR$ and consider $\bR \times [0,1] \times X$. We start from the above map to $\{0\} \times [0,1] \times X$, and slightly perturb it by introducing a function 
\beq
h : L \to \bR.
\eeq
Near a point $p \in S \subset L$ around which the map to $[0,1]\times X$ behaves as \eqref{eq:Jsing}, we take $h$ to be
\beq
h =  \xi \rho( \vec z ) ,
\eeq
where $\rho(\vec z)$ is a function with the property that 
\beq
\rho(\vec z)  = z_n \qquad \text{for} \quad |\vec z| \ll 1,
\eeq
and $\xi \ll 1$ is a sufficiently small constant so that, away from $\vec z=0$, the rank of $\d g$ is not changed by the perturbation by $h$. After this perturbation, we see that $\d g$ always has the maximum rank and hence $g$ is an immersion.
 
The normal bundle $\nu(L)$ is well-defined since $g$ is an immersion. For later purposes, we want to study its Euler class. In general, the Euler class of a vector bundle can be studied by considering a section of the bundle.
On $\bR \times [0,1] \times X$, let us take a vector field $\vec n = (1,0,\cdots,0)^{\rm T}$ which is pointing in the $\bR$ direction. Then we pull back it to $L$, and project it to the normal bundle $\nu(L)$. In this way, we get a section $\vec s$ of $\nu(L)$. As far as the perturbation $h$ is small, the section $\vec s$ is nonzero almost everywhere since $\vec n$ is almost orthogonal to the tangent bundle $T L$. However, $\vec s$ becomes zero at points in $S \subset L$ where the map $g$ before perturbation is singular as in \eqref{eq:Jsing}. A point $p \in S$ contributes $\pm 1$ to the Euler number of $\nu(L)$. We have seen that the number of points in $S$ is even as long as $N_0$ and $N_1$ are embeddings. Thus the Euler number is even. This fact will play an important role in the study of the perturbative anomaly of fermions.

\section{Constraints on Type~IIB string theory}\label{sec:IIBconstraint}

Now we apply the general discussions of Section~\ref{sec:anomalycohomology} to the D3-brane of Type IIB~string theory.

\subsection{Perturbative anomalies}\label{sec:perturbative}

In this subsection, we focus on perturbative anomalies. 
First let us recapitulate what we found in Section~\ref{sec:Homsub}. Any homology class of $ H_5(X; \wt \bZ) $ is realized by a submanifold $N \subset X$. 
Suppose we have two different submanifolds $N_0$ and $N_1$ that represent the same element of the bordism group $\Omega^{\rm SO}_{5, \zeta}(X)$. Let $\nu(N_0)$ and $\nu(N_1)$ be the normal bundles of these submanifolds in $X$. We have found in Section~\ref{sec:betweenN0N1} that there is a bordims between them,
\beq
g : L \to \bR \times [0,1 ] \times X
\eeq
with the following properties.
\begin{itemize}

\item
The boundaries of $L$ are $\partial L = N_1 \sqcup \overline{N_0} $ .

\item Near $\partial L$, the map $g$ is an embedding, $g(N_i) \subset \{0\} \times \{i\} \times X$, and $L$ is transverse to the boundaries $\bR \times \{i\} \times X$.

\item The $g$ is an immersion and hence the normal bundle $\nu(L)$ to $L$ in $ \bR \times [0,1 ] \times X$ is well-defined. On the boundary $N_i$, the $\nu(L)$ becomes $\nu(L)|_{N_i} = \nu(N_i) \oplus \underline{\bR}$, where $\underline{\bR}$ is the trivial bundle. 

\item There is a section $\vec s$ of $\nu(L)$ with the following properties. Near the boundaries $N_i$, the $\vec s$ points in the direction $\underline{\bR}$ of $\nu(L)|_{N_i} = \nu(N_i) \oplus \underline{\bR}$. Also, the $\vec s$ has isolated generic zeros at points $p \in S$, where $S \subset L$ is a set of points with an even number of elements.

\end{itemize}
From the general discussions of anomalies around \eqref{eq:compatibility}, we expect that there is some $6$-form $\sA_6$ such that
\beq
\cZ_\textrm{anomaly}(N_1) \cZ_\textrm{anomaly}(N_0)^{-1} = \exp\left( 2\pi \i \int_{L} \sA_{6} \right). \label{eq:pertanom3}
\eeq
Our purpose is to determine $ \sA_{6}$.

First, let us consider the perturbative anomaly of the fermions. The normal bundle of a D3-brane worldvolume has rank 6, and the Lie algebra of the structure group is $\so(6) = \su(4)$. The D3-brane worldvolume fermions are Weyl fermions $\psi$ in one of the two spinor representations of $\so(6)$, or equivalently the fundamental representation of $\su(4)$. 

Recall that a worldvolume of a D3-brane is possibly a non-orientable manifold. The orientation of the total space $X$ gives the orientation of the sum of the tangent bundle and the normal bundle, and hence the orientation reversal of the tangent and normal bundles are correlated.
The orientation reversal symmetry of a D3-brane acts as a CP symmetry $\psi \leftrightarrow \overline{\psi}$ on the worldvolume fermions, and in particular it acts on $\so(6) = \su(4)$ as the outer automorphism, exchanging the two spinor representations of $\so(6)$ (or equivalently the fundamental and the antifundamental representations of $\su(4)$).

The anomaly polynomial of the fermions is given by 
\beq
\sA_{6}^\textrm{fermion}={\rm ch}_3 = \frac{e}{2},
\eeq
where ${\rm ch}_3$ is the third Chern character of $\su(4)$, and $e$ is the Euler class of $\so(6)$. Under the outer automorphism of $\su(4) = \so(6)$, the ${\rm ch}_3$ and $e$ change sign. This is the appropriate sign change such that the integral $\int_L \sA_{6}^\textrm{fermion} $ is well-defined. 

In general, the Euler class measures the obstruction to the existence of a everywhere nonzero section.
In particular, we have a section $\vec s$ of $\nu(L)$ mentioned above. Without changing the behavior near the boundaries $\partial L =N_1 \sqcup \overline{N_0}$, we may deform the $\so(6)$ connection in such a way that the Euler class is given by
\beq
e = \sum_{p \in S} (\pm 1) \cdot \delta_p,
\eeq
where $\delta_p$ is the delta function 6-form supported on $p \in S$, and the sign $(\pm 1)$ is determined by the behavior of the section $\vec s$ near $p$. Therefore, we obtain
\beq
\int_{L}  \sA_{6}^\textrm{fermion} = \frac12 \sum_{p \in S} (\pm 1 ) . \label{eq:contributionEuler}
\eeq
As mentioned above, the number of points in $S$ is even if $N_0$ and $N_1$ are embeddings. Therefore, from \eqref{eq:pertanom3}, the fermion anomaly $\cZ_{\rm anomaly}^{\rm fermion}$ satisfies
\beq
\cZ_{\rm anomaly}^{\rm fermion}(N_0) = \cZ_{\rm anomaly}^{\rm fermion}(N_1) \quad \text{if $N_0$ and $N_1$ are embeddings.}
\eeq
The reason that we emphasize that $N_0$ and $N_1$ are embeddings is that this equation can be violated if one of $N_i$'s is only an immersion. We will discuss a simple example in Section~\ref{sec:notembedding}. 

We conclude that we can actually take $ \sA_{6}^\textrm{fermion} $ to be zero since it has no effect on \eqref{eq:pertanom3}. This is an important consistency check. 

Next, let us study the perturbative anomaly of the $\U(1)$ gauge field, or the Maxwell field, which we denote as $A$. Let $B_2$ and $C_2$ be the NSNS $B$-field and the 2-form RR field on $X$, respectively. For simplicity, we consider the case that the $\GL(2,\bZ)$ bundle is trivial (see  \cite{Hsieh:2019iba,Hsieh:2020jpj} for the case of $\SL(2,\bZ)$ bundles), but the final result makes sense for more general $\GL(2,\bZ)$ bundles. 

The Maxwell action coupled to $B_2$ and $C_2$ is schematically given by
\beq
 \int_M \left( - \frac{2\pi}{2g_s} (\d A +B'_2) \wedge \star (\d A+B'_2) +\frac{2\pi \i \,C_0 }{2} (\d A +B'_2)^2+ 2\pi \i\, C'_2  (\d A + B'_2) \right) \label{eq:Maxwell}
\eeq
where $g_s$ and $C_0$ are the dilaton and the RR 0-form field, and $B'_2$ and $C'_2$ are roughly given by
\beq
B'_2 = B_2 +\frac12 w_2(M), \qquad C'_2 = C_2 +\frac12 w_2(M), \label{eq:BCprime}
\eeq
where $w_2(M)$ is the second Stiefel-Whitney class of the worldvolume.
The reason for the contribution $\frac12 w_2(M)$ is as follows. We have mentioned around $\eqref{eq:spinc}$ that the $\U(1)$ field on a D-brane satisfies the shifted quantization condition determined by $w_2(M)$. In terms of the $A$ in the above action, the $F$ in $\eqref{eq:spinc}$ is given roughly by $F = \d A + \frac12 w_2(M)$. (See \cite{Hsieh:2020jpj} for more discussions.) We have included this shift $\frac12 w_2(M)$ into $B'$, and regard $A$ as an ordinary $\U(1)$ gauge field rather than a $\spin^c$ connection. The $\frac12 w_2(M)$ in $C'_2$ is determined by the S-duality. The term $\frac12 w_2(M)$ in $B'$ and $C'$ is not necessary for the determination of the perturbative anomaly, but we will use them for our later computation of a global anomaly in Section~\ref{sec:Ktheory} and also in Section~\ref{sec:maxwell}.

The action \eqref{eq:Maxwell} is gauge invariant under gauge transformations of $B_2$ given by $\delta B_2 = \d b_1$ and $\delta A = - b_1$. However, 
it is not invariant under gauge transformations of $C_2$ (and $C_0$). If we perform a gauge transformation $\delta C_2 = \d c_1$, the action changes by an amount
\beq
 \int_M \left( 2\pi \i\, c_1  \d B'_2 \right) .
\eeq
This is cancelled by anomaly inflow as follows. We introduce a manifold $N'$ whose boundary is $\partial N'=M$. We take a term in $N'$ given roughly by
\beq
-2\pi \i \int_{N'}   C'_2 \d B'_2.  \label{eq:CdBanomaly}
\eeq
(More precise meaning of it may be given in terms of differential cochains as discussed in \cite{Hsieh:2020jpj}.)
The variation of the Maxwell action under gauge transformations of $C_2$ is cancelled by the variation of the term in $N'$.\footnote{The variation under $C_0 \to C_0+1$, which is part of the $\GL(2,\bZ)$ transformation, is not yet completely cancelled. Thus there is an anomaly under this transformation which is not captured by \eqref{eq:CdBanomaly}. This will be important in Section~\ref{sec:trivialhomology}, but it does not contribute to the perturbative anomaly.}
The anomaly $\cZ_{\rm anomaly}^{\rm Maxwell}(N)$ of the Maxwell theory for the trivial $\GL(2,\bZ)$ bundle is therefore given by
\beq
\cZ_{\rm anomaly}^{\rm Maxwell}(N) = \exp\left(   -2\pi \i \int_{N}   C'_2 \d B'_2 \right). \label{eq:BCanomaly}
\eeq
This gives not just the perturbative anomaly but also the nonperturbative anomaly (for the trivial $\GL(2,\bZ)$ bundle.)

For the computation of the anomaly polynomial $\sA^{\rm Maxwell}_6$, we only need differential forms and hence we neglect $w_2(M)$. The field strength of $B_2$ and $C_2$ are denoted as
\beq
H_3 = \d B_2, \qquad G_3 = \d C_2.
\eeq
Then it is clear from the defining property \eqref{eq:compatibility} of the anomaly polynomial that $\sA^{\rm Maxwell}_6$ is given by
\beq
\sA^{\rm Maxwell}_6 = -G_3 \wedge H_3.
\eeq
The product $G_3 \wedge H_3$ has the correct $\GL(2,\bZ)$ transformation law and hence this result makes sense even for general $\GL(2,\bZ)$ bundles. 

Combining the Maxwell and the fermion contributions, we finally get the anomaly polynomial of the perturbative anomaly as
\beq
\sA_6 =  -G_3 \wedge H_3. \label{eq:IIBI6}
\eeq
This is really the desired result by the following reason. The modified Bianchi identity \eqref{eq:Bianchi} gives 
\beq
\d G_5 = G_3 \wedge H_3.
\eeq
This is the well-known equation for Type~IIB supergravity.

\subsection{Definition of the cohomology element for the obstruction} \label{sec:Def}
The discussions of Section~\ref{sec:anomalycohomology} have been done without taking into account the fact that $\cZ_{\rm anomaly}(N)$ for the D3-brane is defined only for submanifolds $N \subset X$. Now we remove this restriction and define a ``5-form field'' (or more precisely an element of the differential cohomology group $\widehat{H}^6(X; \wt \bZ)$) associated to the anomaly.

Let $ C \in Z_5(X;\wt \bZ)$ be an arbitrary cycle, where $Z_i(X;\wt \bZ)$ is the set of cycles. We define $\cZ_{\rm anomaly}(C)$ as follows. We take a submanifold $N$ that is homologous to $C$. (More precisely, we take an embedding $ f: N \to X$ such that $[C]=f_* [N]$ where $[N]$ is the fundamental class and $[C]$ is the homology class represented by $C$. We omit this kind of precise statements below.) Such an $N$ is guaranteed to exist by the results of Section~\ref{sec:Homsub}. We also take a chain $D \in C_6(X;\wt \bZ)$, where $C_i(X;\wt \bZ)$ is the set of chains, such that~
\beq
\partial D = N-C.
\eeq
Then we define $\cZ_{\rm anomaly}(C)$ as
\beq
\cZ_{\rm anomaly}(C) : = \cZ_{\rm anomaly}(N) \exp\left( -2\pi \i \int_D \sA_6 \right), \label{eq:defdiffcoh}
\eeq
where $\sA_6$ is given by \eqref{eq:IIBI6}. 

For \eqref{eq:defdiffcoh} to be well-defined, we need to check that it is independent of a choice of $N$ and $D$. There are three steps for this purpose.

\begin{enumerate}

\item
If $D$ is a cycle, i.e. $\partial D =0$, then we have $ \int_D \sA_6 \in \bZ$ if the 3-form field strengths $H_3$ and $G_3$ are such that the de~Rham cohomology class $[\sA_6]$ is in (the image of) integral cohomology. This property guarantees that \eqref{eq:defdiffcoh} is independent of a choice of $D$. 

\item
If there is another submanifold $N_1$ that is bordant to $N_0:=N$, then the result of Section~\ref{sec:perturbative} guarantees that the values of \eqref{eq:defdiffcoh} for $N_0$ and $N_1$ are the same. 

\item
If there is a submanifold $N$ that is nontrivial as an element of the bordism group $\Omega^{\rm SO}_{5, \zeta}(X)$ but is trivial in the homology group $H_5(X; \wt \bZ) $, we have a chain $D$ such that $N = \partial D$ and $D$ cannot be represented by a map from a manifold. We need to have 
\beq
\cZ_{\rm anomaly}(N) =  \exp\left( 2\pi \i \int_D \sA_6 \right)  \label{eq:check}
\eeq
for \eqref{eq:defdiffcoh} to be well-defined. We will show it in Section~\ref{sec:trivialhomology}.

\end{enumerate}

Now from the discussions of Section~\ref{sec:anomalycohomology}, we have a  ``5-form field'' (or more precisely an element of the differential cohomology group $\widehat{H}^6(X; \wt \bZ)$) whose ``field strength 6-form'' is given by $\sA_6 =  -G_3 \wedge H_3$. Topologically, it gives an element 
\beq
\sC_6\in H^6(X ; \wt \bZ), 
\eeq
such that its image $[\sC_6]_\bR$ in the cohomology with real coefficient $H^6(X ; \wt \bR)$ is given by the de~Rham cohomology class $-[G_3 \wedge H_3]$. The de~Rham cohomology describes only partial information. The torsion part corresponds to nonperturbative, global anomalies. In particular, when $\sA_6=0$, the $\sC_6$ comes from the Bockstein homomorphism of the element of $H^5(X; \wt \bR/\bZ)$ determined by $\cZ_{\rm anomaly}$.

The $\sC_6$ is the obstruction to the existence of a shifted quantization. For Type~IIB string theory to make sense, we need $\sC_6=0$ or otherwise there must be D3-branes in a Poincare dual of $\sC_6$ to cancel it.

\subsection{Nonperturbative anomalies}\label{sec:global}

Nonperturbative anomalies are quite complicated and we do not try to give a systematic discussion. We only mention some of the global anomalies and discuss some of them as examples in Section~\ref{sec:ex2}. We only discuss the case $H_3=G_3=0$ so that the perturbative anomaly is absent.

\begin{itemize}

\item Suppose that the $\GL(2,\bZ)$ bundle is trivial. The precise structure group of a 5-manifold in this case is $[\Spin(5) \times \Spin(5)]/\bZ_2$, where the first and the second $\Spin(5)$ are for the tangent and the normal directions of $N$ on which we evaluate the anomaly. Recall that $\Spin(5) = \Sp(2)$. The structure on manifolds with this structure group may be called the $\spin\text{-}\Sp(2)$ structure. (This is a kind of generalization of $\spin$ and $\spin^c$ structures.) Both the fermions and the Maxwell field have global anomalies for this structure~\cite{Witten:1982fp,Wang:2018qoy}.

\item Suppose that the normal bundle is trivial, but we now turn on $\SL(2,\bZ)$ bundles. The spin cover of this group is denoted as $\Mp(2,\bZ)$. The structure group of a 5-manifold is now $[\Spin(5) \times \Mp(2,\bZ) ]/\bZ_2$ and this structure may be called the $\spin\text{-}\Mp(2,\bZ)$ structure. On the fermions, it acts as follows. The abelianization of $\SL(2,\bZ)$ and $\Mp(2,\bZ)$ are $\bZ_{12}$ and $\bZ_{24}$, respectively. Then we obtain a $\spin\text{-}\bZ_{24}$ structure from a $\spin\text{-}\Mp(2,\bZ)$ structure. The $\bZ_{24}$ acts as a symmetry on the fermions, and the element $e^{2\pi \i \, \frac{12}{24} } \in \bZ_{24}$ is identified with the fermion parity $(-1)^{\sF} \in \Spin(5)$. In general, fermions have anomalies under such a discrete symmetry~\cite{Hsieh:2018ifc,Garcia-Etxebarria:2018ajm}. The Maxwell theory also has anomalies~\cite{Seiberg:2018ntt,Hsieh:2019iba,Hsieh:2020jpj}. 

\item  Even if $H_3=G_3=0$, the anomaly \eqref{eq:BCanomaly} can be nonzero. The fields $B_2$ and $C_2$ may be such that their cohomology classes are torsion. Also, $\frac12 w_2(M)$ in \eqref{eq:BCprime} contributes to $B'_2$ and $C'_2$. For instance, these contributions play an important role on a 5-manifold $\mathbb{PR}^5$ surrounding an O$3$-plane. The values $\int_{\mathbb{RP}^5} G_5 = \pm \frac 14$ for four types of O$3$-planes depend on torsion fluxes of $B'_2$ and $C'_2$~\cite{Witten:1998xy}, and they can be understood in terms of the anomaly $\cZ_\textrm{anomaly}(\mathbb{RP}^5)$ \cite{Hsieh:2019iba,Hsieh:2020jpj}.

\end{itemize}

We note that the shift discussed in the example in Section~\ref{sec:exshift} is due to an $\Sp(2)$ anomaly. The normal bundle is such that the $\Sp(2)$ has one instanton number on $S^4 =\mathbb{HP}^1 \subset \mathbb{HP}^2$. Also, the direction $S^1$ has the periodic spin structure. Then one can compute the Atiyah-Patodi-Singer $\eta$-invariant or the mod-2 index to show that the anomaly is given by $\cZ_{\rm anomaly}^{\rm fermion}(S^1 \times S^4) = (-1)$. This is the reason for the shift of $\int_{S^1 \times S^4} G_5 $. On the other hand, if the spin structure on $S^1$ is anti-periodic, the value of $\cZ_{\rm anomaly}^{\rm fermion}$ is trivial and there is no shift.

\subsection{K-theoretic interpretation and its generalization} \label{sec:Ktheory}
The results of the present paper can be put in the context of K-theoretic interpretation of RR fields and its generalization. (This subsection assumes some familiarity with Section 11 of \cite{Diaconescu:2000wy} and may be skipped on a first reading.) 

Let us consider the case that the $\GL(2,\bZ)$ bundle is trivial. 
Type II RR fields are supposed to be described by K-theory~\cite{Moore:1999gb,Freed:2000tt,Freed:2000ta}. This implies, among other things, the following constraint on the RR 2-form field $C_2$ and the NSNS 2-form field $B_2$. Let $\sG_3 \in H^3(X;\bZ)$ and $\sH_3 \in H^3(X;\bZ)$ be the integral cohomology classes of $C_2$ and $B_2$, respectively. First, if $B_2=0$, the condition for $\sG_3$ to have a K-theory lift is given by
\beq
\widehat{Sq}^3 \sG_3 =0, \label{eq:K-lift}
\eeq
where $\widehat{Sq}^3$ is the integral lift of the Steenrod square $Sq^3$, defined as follows. Let $Sq^i~(i=1,2,\cdots $ be the Steenrod squares. First, we reduce $\sG_3$ to an element of $H^3(X; \bZ_2)$ with $\bZ_2$ coefficient. We denote the reduced element by the same symbol $\sG_3$ by abuse of notation. Then we act $Sq^2$. After that, we operate the Bockstein homomorphism $\beta : H^i(X; \bZ_2) \to H^{i+1}(X;\bZ)$. Thus
\beq
\widehat{Sq}^3 \sG_3 = \beta Sq^2 \sG_3 \in H^6(X;\bZ).
\eeq
The mod-2 reduction of $\beta$ is $Sq^1$, and hence the mod-2 reduction of $\widehat{Sq}^3 \sG_3$ is given by $Sq^1 Sq^2 \sG_3 = Sq^3 \sG_3$, where we have used the Adem relation $Sq^1 Sq^2 = Sq^3$. 

When $B_2$ is turned on, the K-theory is twisted and the above condition on $\sG_3$ is modified as~\cite{Witten:1998cd,Bouwknegt:2000qt,Atiyah:2005gu} 
\beq
\widehat{Sq}^3 \sG_3 + \sH_3 \cup \sG_3 =0.  \label{eq:twistedK-lift}
\eeq
However, a question was raised in \cite{Diaconescu:2000wy} about it. This equation is not invariant under the $\SL(2,\bZ)$ (or more generally $\GL(2,\bZ)$) duality symmetry. Thus, it needs to be modified further.

We claim that \eqref{eq:twistedK-lift} should be replaced by the condition obtained in the present paper,
\beq
\sC_6=0. \label{eq:generalcondition}
\eeq
Moreover, it gives a generalization to the case where the $\GL(2,\bZ)$ bundle is nontrivial. 

Let us examine the relation between \eqref{eq:twistedK-lift} and \eqref{eq:generalcondition} in the case of the trivial $\GL(2,\bZ)$ bundle. In fact, $\widehat{Sq}^3 \sG_3 + \sH_3 \cup \sG_3$ can be interpreted as the cohomology class of a differential cohomology element in $\widehat H^6(X;\bZ)$ as follows. First, $\sH_3 \cup \sG_3$ is the cohomology class of the term which we roughly write as $B_2 \d C_3$. Next, $\beta Sq^2 \sG_3$ is the cohomology class of the element $\frac 12 Sq^2 \sG_3 \in H^5(X; \bR/\bZ)$. By using Wu classes $\nu_i$, we can simplify this element as follows. Let $N$ be a closed oriented 5-manifold. We have
\beq
\int_N Sq^2 \sG_3 = \int_N \nu_2(N) \sG_3 = \int_N w_2(N) \sG_3,
\eeq
where we have used the definition of the Wu class $Sq^2 \sG_3|_N = \nu_2(N) \sG_3|_N$, and the fact that the total Wu class $\nu$ and the Stiefel-Whitney class $w$ of a manifold $N$ is related by $w(N) = Sq ( \nu(N) )$, where $Sq$ is the total Steenrod square. Therefore, the differential cohomology element for $\widehat{Sq}^3 \sG_3 + \sH_3 \cup \sG_3$ has the property that its evaluation on a closed oriented manifold $N$ is given by
\beq
\int_N \left( B_2 \d C_2 + \frac12 w_2(N) \sG_3 \right). \label{eq:Kdiff}
\eeq
On the other hand, the Maxwell anomaly \eqref{eq:BCanomaly} is given by
\beq
\frac{1}{2\pi \i} \log \cZ_{\rm anomaly}^{\rm Maxwell}(N)  = \int_N \left( B_2 \d C_2 + \frac12 w_2(N) \sG_3 +  \frac12 w_2(N) \sH_3 + \frac12 w_2(N)w_3(N) \right),
\eeq
where $w_3(N) = Sq^1 w_2(N)$. We see that the Maxwell anomaly contains the terms in \eqref{eq:Kdiff}, as well as correction terms. 

There are ``constant terms'' that are independent of $B_2$ and $C_2$. One constant term comes from the term $ \frac12 w_2(N)w_3(N) $ in the Maxwell anomaly, and the other constant term comes from the fermion anomaly $\cZ_{\rm anomaly}^{\rm fermion}(N)  $. 

We emphasize again that \eqref{eq:generalcondition} is also applicable to the case of nontrivial $\GL(2,\bZ)$ bundles. This is our answer to the question raised in \cite{Diaconescu:2000wy}.

\section{More examples}\label{sec:ex2}

We would like to discuss a few more important examples.

\subsection{Maxwell anomaly}\label{sec:maxwell}
The $\U(1)$ gauge field, which we call the Maxwell theory, on a D3-brane can have global anomalies. We take the $\GL(2,\bZ)$ bundle to be trivial. We also set $B_2=C_2=0$. Even then, the $B'_2$ and $C'_2$ defined in \eqref{eq:BCprime} are nonzero and given by $\frac12 w_2(N)$ which is regarded as an element of $H^2(N;   {\bR/\bZ}  ) $. We denote it as $ w'_2(N)$,
\beq
w'_2(N) = \frac12 w_2(N) \in H^2(N;   {\bR/\bZ}  ).
\eeq
In this case, the anomaly \eqref{eq:BCanomaly} is given as follows. We consider the exact sequence
\beq
\cdots \to  H^2(N;  \bR ) \xrightarrow{ \alpha} H^2(N;   {\bR/\bZ}  ) \xrightarrow{ \beta} H^{3}(N; \bZ ) \to \cdots \label{eq:longexact2}
\eeq
and define
\beq
 W_3(N) =  \beta(w'_2(N)). \label{eq:W3def}
\eeq
The anomaly is then
\beq
\cZ_{\rm anomaly}^{\rm Maxwell}(N) = \exp\left(   -2\pi \i \int_{N}   w'_2(N)  W_3(N) \right)= \exp\left(   -\pi \i \int_{N}   w_2(N)  W_3(N) \right).\label{eq:w2w3anomaly}
\eeq

An example in which the value of this anomaly is nontrivial is given by a $\mathbb{CP}^2$ bundle over $S^1$ constructed as follows~\cite{Freed:1999vc,Wang:2018qoy}. Let $[z_1,z_2,z_3]$ be the homogeneous coordinates of $\mathbb{CP}^2$, i.e. $[z_1,z_2,z_3] $ is the equivalence class of $(z_1, z_2,z_3)$ with the equivalence relation $(\alpha z_1, \alpha z_2, \alpha z_3) \sim (z_1, z_2, z_3)$ for nonzero $\alpha \in \bC$. There is a map $h : \mathbb{CP}^2 \to \mathbb{CP}^2$ given by
\beq
h: [z_1,z_2,z_3] \mapsto [z_1^*,z_2^*,z_3^*] \label{eq:ccmap}
\eeq
which takes complex conjugates of the homogeneous coordinates. We construct a bundle by starting from $[0,1] \times \mathbb{CP}^2$ and then gluing the two ends $\{0\} \times  \mathbb{CP}^2$ and $\{1\} \times  \mathbb{CP}^2$ by using the above map $h$. In this way we obtain a 5-manifold $Z$ which is a bundle of the form
\beq
  \xymatrix{
     \mathbb{CP}^2 \ar[r] & Z \ar[d] \\
    &  S^1
  } \label{eq:CP2S1}
\eeq

Let us recall some basic facts.
On $\mathbb{CP}^2$, we have the canonical complex line bundle, which we denote as $\cL^{-1}$, whose fiber at $[z_1,z_2,z_3]$ is given the line
$ (uz_1, uz_2, uz_3) $ spanned by the vector $(z_1,z_2,z_3)$ where $u \in \bC$. We denote its dual bundle as $\cL$. The $\cL$ is described by equivalence classes $[z_1,z_2, z_3, v]$ of the coordinates $(z_1, z_2, z_3, v)$, where $[z_1,z_2,z_3]$ are the homogeneous coordinates and $v \in \bC$, and the equivalence relation is given by $(z_1, z_2, z_3, v) \sim \alpha (z_1, z_2, z_3, v)$. The pairing between $ (u z_1, u z_2, u z_3) \in \cL^{-1}$ and $[z_1,z_2, z_3, v] \in \cL$ is given by $u v$.
By taking a hermitian metric on $\cL$ (or on $\cL^{-1}$), we can regard the dual line bundles $\cL$ and $\cL^{-1}$ as complex conjugate bundles of each other. The first Chern class $c_1(\cL) \in H^2(\mathbb{CP}^2; \bZ) \simeq \bZ$ is the generator of the cohomology ring $H^\bullet(\mathbb{CP}^2; \bZ)$. 

The complex conjugation map \eqref{eq:ccmap} is naturally lifted to an antilinear bundle map $\cL_{p} \to \cL_{p^*} $ where $p=[z_1, z_2, z_3]$ and $p^*=[z^*_1, z^*_2, z^*_3]$. The antilinear map is equivalent to the linear map $\cL_{p} \to \cL^*_{p^*} $. Then, under the pullback by $h$, the cohomology class $c_1(\cL) \in H^2(\mathbb{CP}^2; \bZ)$ changes the sign as, 
\beq
h^* c_1(\cL) = c_1(\cL^*) = - c_1(\cL).
\eeq
This means that $c_1(\cL)$ cannot be realized as an integeral cohomology class in the 5-manifold $Z$. However, the underlying real vector bundle $\cL_{\bR}$ of $\cL$ can be uplifted to the 5-manifold $Z$ as a non-orientable real rank 2 bundle. By abuse of notation we denote the real bundle on $Z$ also by $\cL_{\bR}$. The mod-2 reduction of $c_1(\cL)$ is the second Stiefel-Whitney class $w_2(\cL_\bR) $, which makes sense even on $Z$. 

We define $ w'_2(\cL_\bR) $ by 
\beq
w'_2(\cL_\bR) = \frac12 w_2(\cL_\bR)  \in H^2(Z; \bR/\bZ).
\eeq
The fact that the integral cohomology class $c_1(\cL) $ is not uplifted to $Z$ means that $ w'_2(\cL_\bR) $ is not contained in the image of $ \alpha$ in
\beq
H^2(Z;  \bR ) \xrightarrow{  \alpha} H^2( Z;   {\bR/\bZ}  )  \xrightarrow{ \beta} H^{3}(Z; \bZ ).
\eeq
Therefore, the element
\beq
\beta ( w'_2(\cL_\bR) ) \in H^{3}(Z; \bZ )
\eeq
is nontrivial. From the above discussions, it is clear that $\beta ( w'_2(\cL_\bR) )  $ is zero if it is restricted to a single fiber $\mathbb{CP}^2$ of \eqref{eq:CP2S1}. Thus it must involve the $S^1$ direction. More precise meaning of this statement is given as follows. Let 
$[\beta ( w'_2(\cL_\bR) )]_2 $
be the mod-2 reduction of $\beta ( \sw'_2(\cL_\bR) )$. Also, let $\sa \in H^1(S^1; \bZ_2)$ be the generator. Then the only possibility compatible with the above discussions is\footnote{In fact, the Wu formula gives $[\beta ( w'_2(\cL_\bR) )]_2 = Sq^1(w_2(\cL_\bR) )=w_1(\cL_\bR) w_2(\cL_\bR)$, and the non-orientability of $\cL_\bR$ along the $S^1$ direction implies that $w_1(\cL_\bR)=\sa$.}
\beq
[\beta (  w'_2(\cL_\bR) ) ]_2 = \sa w_2(\cL_\bR). \label{eq:betaw'2}
\eeq 
By using the fact that $\int_{\mathbb{CP}^2} w_2(\cL_\bR)^2 = 1 \in \bZ_2$, we obtain
\beq
\exp \left( \pi \i \int_Z w_2(\cL_\bR) \beta(w'_2(\cL_\bR)) \right) = -1.
\eeq

We are going to construct a spin 10-manifold by using the 5-manifold $Z$. Before doing that, we recall that $\mathbb{CP}^2$ does not admit a spin structure and $Z$ does not admit a $\spin^c$ structure by the following reason. The tangent bundle $T \mathbb{CP}^2$ is described by infinitesimal deformations $[z_1+\delta z_1, z_2+ \delta z_2, z_3+\delta z_3] $ with the equivalence relation $(\delta z_1, \delta z_2, \delta z_3) \sim (\delta z_1+ \delta \alpha z_1, \delta z_2 + \delta \alpha z_2, \delta z_3 + \delta \alpha z_3)$, where $\delta \alpha \in \bC$ is infinitesimal.
Each infinitesimal deformation $\delta z_i~(i=1,2,3)$ may be regarded as an element of $\cL$ because $(z_1,z_2,z_3, \delta z_1, \delta z_2, \delta z_3)$ and $ ( \alpha z_1,\alpha z_2,\alpha z_3, \alpha \delta z_1, \alpha \delta z_2, \alpha \delta z_3)$ are equivalent for nonzero $\alpha \in \bC$. On the other hand, $[z_1+\delta \alpha z_1, z_2+ \delta \alpha  z_2, z_3+ \delta \alpha z_3]$ may be regarded as the trivial bundle because $\delta \alpha \in \bC$ is just a number. Thus we have
\beq
T \mathbb{CP}^2 \oplus \underline{ \bC} = \cL \oplus  \cL \oplus  \cL , \label{eq:CPtangent}
\eeq
where $ \underline{ \bC} $ is the trivial bundle. From it, the total Chern class of $T \mathbb{CP}^2 $ is determined to be
\beq
c(T \mathbb{CP}^2 ) = ( 1+ c_1(\cL))^3. \label{eq:CPChern}
\eeq
The total Stiefel-Whitney class of $\mathbb{CP}^2$ is the mod-2 reduction of $c(T \mathbb{CP}^2 ) $ as we mentioned around \eqref{eq:CWS}. Thus, $w_2( \mathbb{CP}^2 ) = c_1(\cL) \mod 2$ is nonzero and hence $ \mathbb{CP}^2 $ does not admit a spin structure.

To extend this result to $Z$, we note the following point. The trivial bundle $ \underline{ \bC}$ in \eqref{eq:CPtangent} becomes nontrivial on $Z$, because it is complex-conjugated when we go around $S^1$. Thus it is given by $\underline{\bR} \oplus \widehat{\underline{\bR}}$, where $\widehat{\underline{\bR}}$ is the non-orientable bundle on $S^1$, and $\underline{\bR}$ is the trivial bundle. Also, the real bundle $\cL_\bR$ is not orientable along $S^1$ and hence its first Stiefel-Whitney class is nontrivial. We have
\beq
w_1(\widehat{\underline{\bR}}) = w_1(\cL_\bR) = \sa 
\eeq
where $\sa $ is (the pullback of) the nontrivial element of $H^1(S^1;\bZ_2)$. Noting that the tangent bundle of $S^1$ is trivial, we have
\beq
TZ \oplus \widehat{\underline{\bR}} = \cL_\bR \oplus  \cL_\bR \oplus  \cL_\bR.
\eeq
Therefore, the total Stiefel-Whitney class of $Z$ is given by
\beq
w(TZ) = \frac{(1+\sa + w_2(\cL_\bR))^3}{(1+ \sa)} = 1+ w_2(\cL_\bR) + \sa w_2(\cL_\bR) + w_2(\cL_\bR)^2 .
\eeq
The fact that $w_2(TZ) = w_2(\cL_\bR) \neq 0$ implies that $Z$ does not admit a spin structure. The fact that $w_3(TZ) = \sa w_2(\cL_\bR) \neq 0$ implies that $Z$ does not admit even a $\spin^c$ structure.\footnote{The Wu formula and the fact that $w_1(TZ)=0$ gives $Sq^1 (w_2(TZ) )= w_3(TZ)$. Since $Sq^1$ is the mod-2 reduction of the Bockstein homomorphism, the $w_3$ is the mod-2 reduction of the $W_3$ defined in \eqref{eq:W3def}. The $W_3$ is the obstruction to the existence of a $\spin^c$ structure. \label{eqW3w3}}

To construct a spin manifold, we consider a vector bundle $E$ over $Z$ with the property that
\beq
w(E) = 1+w_2(\cL_\bR) + \sa w_2(\cL_\bR) + \cdots. \label{eq:conditiononE}
\eeq
The total space $E$ has the property that its tangent bundle $T E$ is given by (the pullback of) $TZ \oplus E$, and hence
\beq
w_2(T E) = w_2(TZ) + w_2( E) =0.
\eeq
Therefore, the total space $E$ admits a spin structure.\footnote{If one wishes, one may compactify each fiber $\bR^{{\rm rank }\, E}$ to $S^{\,{\rm rank }\, E}$ to get a closed manifold. But this is not essential at all for our purposes.}

In this subsection, we take as an example
\beq
E = \cL_{\bR} \oplus \widehat{\underline{\bR}}.
\eeq
This bundle satisfies the above condition \eqref{eq:conditiononE}. The total space $E$ is an 8-manifold. (In Section~\ref{sec:trivialhomology} we will consider another example.) The tangent bundle of $E$ is 
\beq
TE =  \cL_\bR \oplus  \cL_\bR \oplus  \cL_\bR \oplus  \cL_\bR \label{eq:fourLR}
\eeq

First let us take the 10-manifold $X$ to be $X = \bR^2 \times E$ and study the shifted quantization condition. We take the submanifold $N$ to be 
\beq
N = \{ (0,0) \} \times Z \subset \bR^2 \times E,
\eeq
where $Z$ is identified as the zero section of $E$. The Maxwell anomaly \eqref{eq:w2w3anomaly} is given by
\beq
\cZ_{\rm anomaly}^{\rm Maxwell}(N) =  \exp\left(   -\pi \i \int_{Z}   w_2(Z)  W_3(Z) \right) =-1, \label{eq:CPmaxanom}
\eeq
where we have used the fact that the mod-2 reduction of $ W_3(Z) $ is given by $ \sa w_2(\cL_\bR)$, which follows from $w_2(Z) =w_2(\cL_\bR)$ and \eqref{eq:betaw'2}. Alternatively, one can use the fact that $w_3(Z)$ is the mod-2 reduction of $W_3(Z)$ (see footnote \ref{eqW3w3}).
Therefore the anomaly is nontrivial. 

For the fermions, we proceed as follows. (Our discussions will be brief. See \cite{Wang:2018qoy} for details). The anomaly of the fermions coupled to a $\spin\text{-}\Sp(2)$ structure is determined by counting the number of zero modes modulo 2 (i.e. the mod-2 index) on $N \simeq Z$. The $Z$ is a fiber bundle with the fiber $\mathbb{CP}^2$ and the base $S^1$, so we first study zero modes on $\mathbb{CP}^2$ and perform dimensional reduction. The normal bundle $\nu(Z)$ in $X$ restricted to a fiber $\mathbb{CP}^2 \subset Z$ is just $\cL_\bR \oplus \underline{\bR}^3$. Its associated ``spin bundle'' is $2(\cL^{1/2} \oplus \cL^{-1/2})$, where $\cL^{1/2}$ should be interpreted as a $\spin^c$ structure for $\mathbb{CP}^2$. 
By using the Atiyah-Singer index theorem, one finds that there is no zero mode on $\mathbb{CP}^2$. (The first Pontryagin class of $\mathbb{CP}^2$ is $p_1(\mathbb{CP}^2) = 3 c_1(\cL)^2$ as can be seen from \eqref{eq:CPChern}.) Therefore, there is no zero mode also on $Z$ and the mod-2 index is zero. We conclude that the fermion contribution is trivial, $\cZ_{\rm anomaly}^{\rm fermion}(N) =1$.

We found that for $N= \{ (0,0) \} \times Z$ the anomaly is given by
\beq
\cZ_{\rm anomaly}(N) =-1.
\eeq
Therefore, the shifted quantization condition is
\beq
\int_N G_5 \in \frac12 + \bZ.\label{eq:shiftbyMaxwell}
\eeq

To construct an example in which the shifted quantization is obstructed (i.e. $\sC_6$ is nonzero), we can simply follow the same strategy as in Section~\ref{sec:Mobs}. We consider the 10-manifold given by $X = \bR \times \widetilde S^1 \times E$, and include a nontrivial holonomy of an orientation reversal element of $\GL(2,\bZ)$ such as $\Omega \in \GL(2,\bZ)$ around $\widetilde S^1$. The $\Omega$ changes the sign of $G_5$ as $G_5 \to -G_5$. Thus \eqref{eq:shiftbyMaxwell} is inconsistent unless we include odd number of D3-branes.

\paragraph{A purely gravitational example.}

We may also construct a ``purely gravitational'' example of nonzero $\sC_6$ in which the $\GL(2,\bZ)$ bundle is trivial. It is given by the total space of a bundle
\beq
  \xymatrix{
     E \times S^1_B \ar[r] & X \ar[d] \\
    &  \wt S^1
  } \label{eq:puregravity}
\eeq
The bundle structure is as follows. The 5-manifold $Z$ is constructed as a bundle over $S^1$, and we denote it as $S^1_A$ to distinguish several $S^1$'s. When we go around the base $\wt S^1$ of \eqref{eq:puregravity}, we flip the directions of both $S^1_A$ and $S^1_B$. In more detail, the action on $Z$ and $E$ is as follows. We can construct $Z$ from $[-1 ,1] \times \mathbb{CP}^2$ by identifying the two boundaries using the map $h$ defined in \eqref{eq:ccmap}. When we go around $\wt S^1$, a point $ (t, [z_1,z_2,z_3]) \in [-1,1] \times \mathbb{CP}^2$ is mapped to $ (-t, [z_1,z_2,z_3])$. In particular, the point $(1, [z_1,z_2,z_3])$ is mapped to $(-1, [z_1,z_2,z_3])$ which is the same as $(1, [z^*_1,z^*_2,z^*_3])$. The action on the 8-manifold $E$ is similar. 

The bundle $\cL_\bR$ can be extended to the 10-manifold $X$, which we denote by the same symbol $\cL_\bR$. Let $\sb \in H^1(X;\bZ)$ be the pullback of the generator of $ H^1(\wt S^1; \bZ_2)$. By the dimensional reason, we have $\sb^2=0$. The total Stiefel-Whitney class of $X$ is given by
\beq
w(X) = (1+ \sb)^2 ( 1+ w_1(\cL_\bR) + w_2(\cL_\bR))^4,
\eeq
where $(1+\sb)^2$ comes from the directions $TS^1_A \oplus T S^1_B$, and $( 1+ w_1(\cL_\bR) + w_2(\cL_\bR))^4$ comes from the directions $T\mathbb{CP}^2 \oplus  \cL_{\bR} \oplus \widehat{\underline{\bR}}$.
In particular we have $w_1(X)=w_2(X)=0$ and hence $X$ admits a spin structure. 

On $X$, we have a submanifold $N \simeq Z$ as before, and the shifted quantization condition \eqref{eq:shiftbyMaxwell} must be satisfied. However, when we gradually move $N$ and go around $\wt S^1$, the orientation of $N$ is reversed. Then, \eqref{eq:shiftbyMaxwell} becomes inconsistent (unless there is a D3-brane). This implies that $\sC_6$ is nonzero.

\subsection{Submanifolds that are homologically trivial}\label{sec:trivialhomology}

For the anomaly to be described by ordinary cohomology (or more precisely its differential version $\widehat H^6(X; \wt \bZ)$), we need to check the property stated around \eqref{eq:check}. The perturbative anomaly is not essential in the following discussion and we set $H_3=G_3=0$. Then we need to show that the anomaly is zero for submanifolds $N \subset X$ whose homology classes in $X$ are trivial. 

 From the discussion of Section~\ref{sec:bortohom}, we essentially need to consider only two cases;
\begin{itemize}
\item $X = \bR^{10}$ and $N = Z$, where $Z$ is the 5-manifold given in \eqref{eq:CP2S1} and is embedded in $\bR^{10}$. 
\item $X = S^1 \times \bR^{9}$ and $N = S^1 \times \mathbb{CP}^2$, where $\mathbb{CP}^2$ is embedded in $ \bR^9$. 
\end{itemize}
The $Z$ represents the generator of $\Omega^{\rm SO}_5({\rm pt}) \simeq \bZ_2 $ since it can be detected by a nonzero Stiefel-Whitney number $\int_Z w_2(X) w_3(Z)$.

\subsubsection{The case $X=\bR^{10}$}
By Whitney embedding theorem, we can embed any $n$-manifold into $ \bR^{2n}$. Thus the $Z$ in \eqref{eq:CP2S1} can be embedded in $X=\bR^{10}$. For our purposes, we need some information of the normal bundle. Thus let us construct an embedding $Z \subset \bR^{10}$ more explicitly (see e.g. \cite{Freed:1999vc}). 

First we construct an embedding of $\mathbb{CP}^2$ in $\bR^8$. Let $[z_1,z_2,z_3]$ be the homogeneous coordinates and we denote $\vec z = (z_1, z_2, z_3)^{\rm T}$ and $[\vec z\,]=[z_1,z_2,z_3]$. Then we consider a function from $\mathbb{CP}^2$ to the space of $3 \times 3$ hermitian traceless matrices given by
\beq
M([\vec z\,]) = \frac{\vec z \cdot \vec z\,^\dagger}{|\vec z|^2} - \frac{1}{3} I_3,
\eeq
where $I_3$ is the unit matrix. As a manifold, the space of $3 \times 3$ hermitian traceless matrices is just $\bR^8$. Thus we obtained a map $\mathbb{CP}^2 \to \bR^8$. The image $M([\vec z\,]) $ uniquely determines $[\vec z\,]$ as the engenspace of the matrix with eigenvalue $2/3$, and hence it is an embedding.  Actually, this is contained in a sphere $S^7 \subset \bR^8$ since 
\beq
\tr M([\vec z\,]) ^2 = \frac23. \label{eq:S7}
\eeq

Next we construct an embedding of $Z$. First, we consider a 10-manifold $Y$ that is a bundle
\beq
  \xymatrix{
     \bR^9 \ar[r] & Y \ar[d] \\
    &  S^1
  } \label{eq:R9S1}
\eeq
with the following bundle structure. We regard $\bR^9$ as $\bR^8 \times \bR$. The part $\bR^8$ is regarded as the space of $3 \times 3$ hermitian traceless matrices, and we define $h' : \bR^8 \to \bR^8$ to be the map that takes complex conjugates of matrices. Also, we define $h'' : \bR \to \bR$ to be the sign change of real numbers. Combining $h'$ and $h''$, we get a map $h = (h', h'')$,
\beq
h: \bR^9 \to \bR^9. \label{eq:compcomj}
\eeq 
This is simply a map which flips the signs of four of the nine coordinates of $\bR^9$. By using $h$ as a transition function, we construct the above bundle $Y$. This bundle is topologically trivial, so $Y $ is just diffeomorphic to $ \bR^9 \times S^1$. However, the representation \eqref{eq:R9S1} will be geometrically more convenient. 

Notice that $\bR^9 \times S^1$ can be embedded in $\bR^{10}$. We first embed $S^1$ in $\bR^{10}$, and then take a tubular neighborhood $S^1 \times D^9$ of it, and finally use a diffeomorphism of $\bR^9$ and an open disk ${\rm int}(D^9)$.

As we have discussed above, the $\mathbb{CP}^2 $ can be embedded in $\bR^8$. By construction, the 5-manifold $Z$ is constructed by using complex conjugation $\vec z \to \vec z\,^*$, so it is clear that $Z$ given by \eqref{eq:CP2S1} can be embedded in $Y$ given by \eqref{eq:R9S1}. Since $Y$ can be embedded in $\bR^{10}$, it is enough to consider the embedding of $Z$ in $Y$. Thus we just consider $Y$ as the spacetime 10-manifold.

The normal bundle is described as follows. First we consider the normal bundle to $\mathbb{CP}^2$ in $\bR^8$. Let $\delta \vec z$ be an infinitesimal variation of $\vec z$ at a point $[\vec z\,]$. The normal bundle is given by the space of $3 \times 3$ hermitian traceless matrices $B$ such that it is orthogonal to the tangent bundle in the sense that
\beq
\tr ( B \delta M([\vec z\,])) =0,
\eeq
where $\delta M([\vec z\,])$ is the infinitesimal change of $M([\vec z\,])$ under the variation $\delta \vec z$. 

An important point about the normal bundle to $\mathbb{CP}^2$ in $\bR^8$ is that the normal bundle is a direct sum of the trivial bundle and a rank 3 bundle, 
\beq
\nu(\mathbb{CP}^2) =   \cN_3 \oplus \underline{\bR} , \label{eq:normalCP2}
\eeq
where $\cN_3$ is a real rank 3 bundle.
This is because we have a everywhere nonzero section $B = M([\vec z\,])$. This is due to the fact that $\mathbb{CP}^2$ is actually embedded in $S^7 \subset \bR^8$ as mentioned around \eqref{eq:S7}, and the normal bundle to $S^7$ in $\bR^8$ is trivial. 

The normal bundle structure \eqref{eq:normalCP2} has the following consequence. Let $\cS_\nu$ be the ``spin bundle'' associated to the normal bundle. (This bundle itself is actually not defined. The ``spin bundle'' associated to the tangent bundle is also not defined since $\mathbb{CP}^2$ is not a spin manifold. However, the tensor product of these two ``spin bundles'' is defined and has a $\spin\text{-}\Sp(2)$ structure.) A priori, the Lie algebra of the structure group of the normal bundle is $\so(4) = \su(2) \times \su(2)$ and the spin bundle is of the form $\cS_\nu = \cS_{\nu, +} \oplus \cS_{\nu,-}$ where $\cS_{\nu, \pm}$ are associated to each of the $\su(2)$'s. The fact that the normal bundle contains the trivial bundle $\underline{\bR}$ as a direct summand implies that only the diagonal part $\su(2) \subset \su(2) \times \su(2)$ is nontrivial, and hence $\cS_{\nu, +} \simeq \cS_{\nu,-}$ which we denote as $\cS_{\nu,0}$,
\beq
\cS_\nu =\cS_{\nu,+} \oplus  \cS_{\nu,-} \simeq \cS_{\nu,0} \oplus  \cS_{\nu,0}.
\eeq

Next we consider the normal bundle to $Z$ in $Y$. It is of the form
\beq
\nu(Z) = \cN_3 \oplus \underline{\bR} \oplus \widehat{ \underline{\bR}},
\eeq
where $\cN_3$ is a rank 3 bundle such that its restriction to a fiber $\mathbb{CP}^2 \subset Z$ is the same as the $\cN_3$ in \eqref{eq:normalCP2}, and $\widehat{ \underline{\bR}}$ is the bundle which flips the sign when we go around the base $S^1$ of $Z$. The part $\widehat{ \underline{\bR}}$ comes directly from the direction $\bR$ of $\bR^{9} = \bR^8 \times \bR$ in the construction of $Y$. 

An important point is that the orientation of $\cN_3$ is changed when we go around the base $S^1$ of $Z$.
The orientation of $\nu(Z)$ is preserved because of the additional bundle $\widehat{ \underline{\bR}}$. This orientation flip of $\cN_3$ implies that when we go around $S^1$, the two bundles $\cS_{\nu, +} $ and $\cS_{\nu, -}$ are  exchanged. 

Now we have enough information to compute the anomalies of the fermions and the Maxwell field. The anomaly of the Maxwell field is completely the same as in Section~\ref{sec:maxwell} and is given by \eqref{eq:CPmaxanom}. Thus we focus on the fermion anomaly.

Let us compute the mod-2 index for the fermions. First we compute the number of zero modes on $\mathbb{CP}^2$. The first Pontryagin class of the tangent bundle $T\mathbb{CP}^2$ is given by 
\beq
p_1(\mathbb{CP}^2) = 3c_1(\cL)^2
\eeq
where $c_1(\cL)$ is the generator of $H^2(\mathbb{CP}^2; \bZ)$ as in Section~\ref{sec:maxwell}. Since the tangent bundle of the 10-manifold $Y$ is trivial, the normal bundle $\nu(\mathbb{CP}^2)$ and hence the bundle $\cN_3$ in \eqref{eq:normalCP2} has the first Pontryagin class $p_1(\cN_3) = -3c_1(\cL)^2$. The corresponding spin bundle $ \cS_{\nu,0}$ has the second Chern character given by
\beq
{\rm ch}_2( \cS_{\nu,0}) = \frac14 p_1(\cN_3)  =  -\frac{3}{4}c_1(\cL)^2.
\eeq
The index ${\rm index}(\mathbb{CP}^2,  \cS_{\nu,\pm})$ of the Dirac operator on $\mathbb{CP}^2$ coupled to the bundle $ \cS_{\nu,\pm}$ is therefore given by the Atiyah-Singer index theorem as
\beq
{\rm index}(\mathbb{CP}^2,  \cS_{\nu,\pm}) &={\rm index}(\mathbb{CP}^2,  \cS_{\nu,0}) \nonumber \\
&= \int_{\mathbb{CP}^2} \left( {\rm ch}_2( \cS_{\nu,0}) - 2 \cdot \frac{p_1(\mathbb{CP}^2) }{24} \right) \nonumber \\
&=-1. \label{eq:ASindex}
\eeq
Thus each $\cS_{\nu,+}$ and $\cS_{\nu,-}$ has a single zero mode, and in total we have two zero modes.

Under the action of $h: \bR^9 \to \bR^9$ in \eqref{eq:compcomj}, the two zero modes are exchanged because $\cS_{\nu,+}$ and $\cS_{\nu,-}$ are exchanged as we mentioned above. Moreover, $h$ has the property that its square is the identity, $h^2=1$. This is true even for the uplift of $h$ to the spin bundle, since $h$ flips four of the nine coordinates of $\bR^9$. From these properties of $h$, we see that one linear combination of the two zero modes survives as a zero mode on the 5-manifold $Z$. The mod-2 index is therefore nonzero, and hence we get the fermion anomaly
\beq
\cZ^{\rm fermion}_{\rm anomaly}(Z)=-1. 
\eeq

The Maxwell and the fermion anomalies cancel with each other and we get
\beq
\cZ_{\rm anomaly}(Z)=1.
\eeq
This is the desired result since the submanifold $Z \subset \bR^{10}$ is homologically trivial. 

\subsubsection{The case $X = S^1 \times \bR^9$}

Next we consider the case that  $X = S^1 \times \bR^{9}$ and $N = S^1 \times \mathbb{CP}^2$, where $\mathbb{CP}^2$ is embedded in $ \bR^9$. 
In this case, it is possible to have a nontrivial $\SL(2,\bZ)$ holonomy around the $S^1$. The holonomy cannot be an orientation-reversing element of $\GL(2,\bZ)$ because in that case the homology class of $N$ is not an element of $H^5(X; \wt \bZ)$ with the correct coefficient system $\wt \bZ$. Thus it is sufficient to consider $\SL(2,\bZ)$.

The anomaly can be detected as follows. Let $\cZ_{\rm matter} (M)$ be the partition function or the path integral measure\,\footnote{
The partition functions of the fermions and the Maxwell field are actually zero in the current situation. For the fermions, it is due to zero modes. For the Maxwell field, it is due to a term $\pi \i \int w_2(M) F$ in the action \eqref{eq:Maxwell}. Because of this term, contributions to the path integral from $F$ and $-F$ have the relative phase difference $\pi \i \int w_2(M)^2$ and they cancel with each other when $M =\mathbb{CP}^2$. Of course, correlation functions can be nonzero. When we say a ``path integral measure'', we include the exponential of the action in the definition of the measure.
} 
of the worldvolume theory of a D3-brane on a worldvolume $M$. For any element $U \in \SL(2,\bZ)$, the $\cZ_{\rm matter}(M)$ transforms as
\beq
\cZ_{\rm matter}(M) \xrightarrow{U} \cZ_{\rm anomaly}(N) \cZ_{\rm matter}(M) ,
\eeq
where $N = S^1 \times M$ with the holonomy $U$ included in the $S^1$ direction. Thus we can study the anomaly $\cZ_{\rm anomaly}(N) $ from the transformation of the $\cZ_{\rm matter}(M)$.

We denote the partition functions (or the path integral measures) of the fermions and the Maxwell field on $\mathbb{CP}^2$ as $\cZ^{\rm fermion}(\mathbb{CP}^2)$ and $\cZ^{\rm Maxwell}(\mathbb{CP}^2)$, respectively.

We first study the transformation of $\cZ^{\rm fermion}(\mathbb{CP}^2)$.
As we have seen around \eqref{eq:ASindex}, there are two zero modes of the fermions on $\mathbb{CP}^2$. These fermions transform under $\bZ_{24}$ which is the abelianization of the double cover ${\rm Mp}(2,\bZ)$ of $\SL(2,\bZ)$. Let $T', S' \in   {\rm Mp}(2,\bZ)$ be the lifts of the usual $T, S \in \SL(2,\bZ)$. They satisfy
\beq
S'^2=(T'^{-1} S')^3, \qquad S'^8=1.
\eeq
(For $T, S \in \SL(2,\bZ)$, we instead have $S^2=(T^{-1} S)^3$ and $S^4=1$. Intuitively, $S$ is a ``$\frac{\pi}{2} $ rotation'' and $T^{-1}S$ is a ``$\frac{\pi}{3}$ rotation''.)
From these relations, one can see that $T'$ corresponds to the generator of $\bZ_{24}$. 
Under this generator, $\cZ^{\rm fermion}(\mathbb{CP}^2)$ transforms due to the axial anomaly as
\beq
\cZ^{\rm fermion}(\mathbb{CP}^2) \xrightarrow{T’} \exp\left( - \frac{2 }{24} \cdot  2\pi \i  \right) \cZ^{\rm fermion}(\mathbb{CP}^2), \label{eq:fS1CP2}
\eeq
where the factor $2$ in the exponent reflects the fact that there are two zero modes on $\mathbb{CP}^2$.\footnote{We have not carefully studied the chirality and the sign of the $\bZ_{24}$ action on fermions. However, we have checked the following. The Maxwell theory can be obtained from a $T^2$ compactification of a chiral 2-form field in 6-dimensions. When the normal bundle is trivial and the manifold is spin, the gravitational anomaly of the chiral 2-form field on an M5-brane is $14$ times the anomaly of the fermions on the same M5-brane because of the transition between the 2-form and $2 \times 14$ fermions via the E-string theory. From this fact, we can fix the sign of the axial anomaly by using the result for the Maxwell theory obtained later. Once we fix the sign, we can apply the result for more general cases of non-spin manifolds with nontrivial normal bundles.}

On the other hand, the anomaly of the Maxwell field under $\SL(2,\bZ)$ transformations is studied in \cite{Witten:1995gf,Seiberg:2018ntt}. Let us regard $\cZ^{\rm Maxwell}(\mathbb{CP}^2)$ as a function of 
\beq
\tau = C_0+ \i g_s^{-1},
\eeq
where $g_s$ is the dilaton (or the gauge coupling squared), and $C_0$ is the RR 0-form field (or the $\theta$-angle). The action is given by \eqref{eq:Maxwell}. Then $\cZ^{\rm Maxwell}(\mathbb{CP}^2)$ transforms as
\beq
\cZ_{\rm bare}^{\rm Maxwell}(\tau+1) &= \exp\left( \frac{2\pi \i }{8}  \sigma \right) \cZ_{\rm bare}^{\rm Maxwell}(\tau) , \label{eq:Ttr} \\ 
\cZ_{\rm bare}^{\rm Maxwell}(-1/\tau) & = \alpha \tau^{(\chi+\sigma)/4}  \overline{\tau}^{(\chi-\sigma)/4} \cZ_{\rm bare}^{\rm Maxwell}(\tau), \label{eq:Str}
\eeq
where $\chi$ and $\sigma$ are the Euler number and the signature of the 4-manifold, respectively, and $\alpha \in \U(1)$ is a phase factor that is independent of $\tau$. The reason for the subscript ``bare'' will become clear soon. The second equation \eqref{eq:Str} was obtained in \cite{Witten:1995gf}, but the phase factor $\alpha$ was not precisely computed for non-spin manifolds. The first equation \eqref{eq:Ttr} is derived as follows. We set $B_2=C_2=0$. From \eqref{eq:Maxwell}, we see that the $C_0$ is coupled to the Maxwell field as
\beq
\frac{2\pi \i}{2} \int C_0 F \wedge F,
\eeq
where $F$ is the field strength of the $\spin^c$ connection which is roughly given by $F= \d A+ \frac12 w_2(N)$. On spin manifolds, we would have $\frac{1}{2} \int  F \wedge F \in \bZ$. However, on non-spin manifolds, this quantization is violated. The Atiyah-Singer index theorem gives
\beq
\frac{1}{2} \int  F \wedge F - \frac{\sigma}{8} \in \bZ,
\eeq
where we have used the signature index theorem $\sigma = \frac13 \int_M p_1(M)$.
Therefore we obtain \eqref{eq:Ttr} under the transformation $C_0 \to C_0+1$. 

The transformation \eqref{eq:Str} contains the factor $\tau^{(\chi+\sigma)/4}  \overline{\tau}^{(\chi-\sigma)/4}$ which is not a $\U(1)$ phase. However, we want anomalies to be pure $\U(1)$ phase factors. To achieve it, we add a counterterm and modify $\cZ^{\rm Maxwell}(\tau) $ as~\cite{Seiberg:2018ntt}
\beq
\cZ^{\rm Maxwell}(\tau) = \eta(\tau)^{-(\chi+\sigma)/2}\eta(- \overline\tau)^{-(\chi-\sigma)/2}\cZ_{\rm bare}^{\rm Maxwell}(\tau),
\eeq
where $\eta(\tau)$ is the Dedekind $\eta$-function. By using $\eta(\tau+1) = e^{\frac{2\pi \i}{24} } \eta(\tau)$ and $\eta(-1/\tau) = \sqrt{-\i \tau}\eta(\tau)$, we get
\beq
\cZ^{\rm Maxwell}(\tau+1) &= \exp\left( \frac{2\pi \i }{12}  \sigma \right) \cZ^{\rm Maxwell}(\tau) , \label{eq:Ttr2} \\ 
\cZ^{\rm Maxwell}(-1/\tau) & = \alpha\exp\left( \frac{2\pi \i }{8}  \sigma \right) \cZ^{\rm Maxwell}(\tau). \label{eq:Str2}
\eeq
Now these anomalies are pure $\U(1)$ phases. 

The signature of $\mathbb{CP}^2$ is given by $\sigma=1$, and hence we obtain
\beq
\cZ^{\rm Maxwell}(\mathbb{CP}^2) \xrightarrow{T’} \exp\left(  \frac{1}{12} \cdot  2\pi \i  \right) \cZ^{\rm Maxwell}(\mathbb{CP}^2). \label{eq:MS1CP2}
\eeq
We see that \eqref{eq:fS1CP2} and \eqref{eq:MS1CP2} exactly cancel with each other. We have not determined the phase $\alpha$, but from the analysis of bordism, it is known~\cite{Hsieh:2019iba,Hsieh:2020jpj} that the anomaly under $S \in \SL(2,\bZ)$ (or more precisely its lift $S' \in {\rm Mp}(2,\bZ)$) is completely fixed by the anomaly under $T'$. Therefore we have demonstrated the complete cancellation of the anomalies of the fermions and the Maxwell field for $N = S^1 \times \mathbb{CP}^2$. 

In summary, we have demonstrated that \eqref{eq:check} holds for elements in the kernel of $\Omega^{\rm SO}_{5, \zeta}(X) \to H_5(X; \wt \bZ)$. This is one of the important ingredients of the construction in Section~\ref{sec:Def}.

\subsection{An immersion that is not an embedding}\label{sec:notembedding}
Here we would like to discuss a subtlety when $f : N \to X$ is not an embedding but just an immersion. In that case, the normal bundle is well-defined and we can compute the anomaly $\cZ_{\rm anomaly}(N)$. However, we do not get the result expected from cohomology. This is the reason that we only used embeddings of $N$ in the construction in Section~\ref{sec:Def}.

Let us just consider the 10-manifold discussed in Section~\ref{sec:exshift}, that is $X = \bR \times S^1 \times \mathbb{HP}^2$ where $S^1$ is a circle. We take $N $ to be $S^1 \times S^4$. The difference from the case of Section~\ref{sec:exshift} is that we consider a map 
\beq
f_2 : S^1 \times S^4 \to \bR \times S^1 \times \mathbb{HP}^2 ,
\eeq
in which the part $S^1 \to S^1$ wraps twice (i.e. the mapping degree is 2). The map $f_2$ can be realized as an immersion, but not as an embedding.

We can define the spin structure of $f_2^* TX$ on $N$ by the pullback of the spin structure on $X $. Regardless of the spin structure on the target space $S^1$ , the spin structure on the worldvolume $S^1$ is periodic. By computing the mod-2 index, we get a nonzero anomaly. To emphasize the fact that the anomaly depends on the map $f_2$, we denote it as $\cZ_{\rm anomaly}(N, f_2)$. We have
\beq
\cZ_{\rm anomaly}(N, f_2)=-1. \label{eq:twicevalue}
\eeq 

Let $ f_1 : S^1 \times S^4 \to \bR \times S^1 \times \mathbb{HP}^2 $ be the map considered in Section~\ref{sec:exshift} in which the part $S^1 \to S^1$ wraps only once. In ordinary homology, we have
\beq
(f_2)_* [N] = 2 (f_1)_* [N] ,
\eeq
where $[N] \in H_5(N; \bZ)$ is the fundamental class of $N$. This is true even in the bordism group $\Omega^{\rm SO}_5(X)$. Let $[ f_i : N \to X] \in \Omega^{\rm SO}_5(X)$ ($i=1,2$) be the classes in the bordism group. Then we have
\beq
[ f_2 : N \to X] = 2 [ f_1 : N \to X].
\eeq
However, since $\cZ_{\rm anomaly}(N, f_2) = -1$ we have
\beq
\cZ_{\rm anomaly}(N, f_2) \neq \cZ_{\rm anomaly}(N, f_1)^2.
\eeq
In fact, the important relation \eqref{eq:pertanom3} with $\sA_6= -G_3 \wedge H_3$ is valid only if both $N_1$ and $N_2$ are embeddings. If one of them is not an embedding, the contribution from the Euler number \eqref{eq:contributionEuler} is not guaranteed to be trivial. 

One can check that if $f_2$ is perturbed in an appropriate way, then there is a single self-intersection point in the immersion $f_2 : N \to X$. On the other hand, two copies of $ f_1 : N \to X$ can be perturbed to be an embedding. A local behavior of the bordism between them may be captured by the bordism \eqref{eq:singlesingular}. There is a single point contributing to the Euler number, and hence \eqref{eq:contributionEuler} leads to
\beq
\cZ_{\rm anomaly}(N, f_2) = - \cZ_{\rm anomaly}(N, f_1)^2.
\eeq
This result is consistent with the value \eqref{eq:twicevalue}. 

Our definition of $\cZ_{\rm anomaly}(C)$ in Section~\ref{sec:Def} have used only embeddings $f : N \to X$ and hence technically there is no problem in our construction. Conceptually, however, it is a curious question why $N$ needs to be an embedding. A D3-brane worldvolume $M$ needs to be an embedding, or otherwise we need to take into account additional light degrees of freedom that appear on intersection points. However, $N$ is not a physical worldvolume of some brane. We postpone more study on this point for future work.

\section*{Acknowledgements}
The work of KY is supported in part by JST FOREST Program (Grant Number JPMJFR2030, Japan), 
MEXT-JSPS Grant-in-Aid for Transformative Research Areas (A) ”Extreme Universe” (No. 21H05188),
and JSPS KAKENHI (17K14265).

\bibliographystyle{ytphys}
\bibliography{ref}

\end{document}